\documentclass[twocolumn]{aastex631}

\usepackage{natbib}

\usepackage{verbatim}
\usepackage{lineno}
\shorttitle{DLA finder for DESI}
\shortauthors{Wang et al. }

\graphicspath{{./}{figures/}}

\begin{document}

\title{Deep Learning of DESI Mock Spectra to Find Damped Ly$\alpha$ Systems}

\correspondingauthor{Zheng Cai}
\email{zcai@mail.tsinghua.edu.cn}

\author[0000-0003-4877-1659]{Ben Wang}
\affiliation{Department of Astronomy,Tsinghua University,Beijing 100084, China }

\author[0000-0001-9189-0368]{Jiaqi Zou}
\affiliation{Department of Astronomy,Tsinghua University,Beijing 100084, China }

\author{Zheng Cai}
\affiliation{Department of Astronomy,Tsinghua University,Beijing 100084, China }

\author{J. Xavier Prochaska}
\affiliation{Department of Astronomy and Astrophysics, UCO,
Lick Observatory, University of California\\
1156 High Street,
Santa Cruz, CA 95064, USA}
\affiliation{Kavli IPMU, the University of Tokyo (WPI), Kashiwa 277-8583, Japan
}

\author{Zechang Sun}
\affiliation{Department of Physics,Tsinghua University,
Beijing 100084, CHINA }

\author{Jiani Ding}
\affiliation{Department of Astronomy and Astrophysics, UCO,
Lick Observatory, University of California\\
1156 High Street,
Santa Cruz, CA 95064, USA}

\author{Andreu Font-Ribera}
\affiliation{Institut de F\'isica d\'Altes Energies (IFAE), 
The Barcelona Institute of Science and Technology, 
08193 Bellaterra (Barcelona), Spain }

\author{Alma Gonzalez}
\affiliation{Consejo Nacional de Ciencia y Tecnología, Av. Insurgentes Sur 1582. Colonia Crédito Constructor, Del. Benito Juárez C.P. 03940, México D.F. México}
\affiliation{Departamento de Física, División de Ciencias e Ingenierías, Campus Leon, Universidad de Guanajuato, León 37150, México}

%í
%\affiliation{Department of Physics and Astronomy,
%University College London, Gower Street\\
%London WC1E 6BT, United Kingdom}

\author{Hiram K. Herrera-Alcantar}
\affiliation{Consejo Nacional de Ciencia y Tecnología, Av. Insurgentes Sur 1582. Colonia Crédito Constructor, Del. Benito Juárez C.P. 03940, México D.F. México}
\affiliation{Departamento de Física, División de Ciencias e Ingenierías, Campus Leon, Universidad de Guanajuato, León 37150, México}

\author{Vid Irsic}
\affiliation{Kavli Institute for Cosmology,
University of Cambridge,
Madingley Road, Cambridge CB3 0HA, UK}
\affiliation{Cavendish Laboratory,
University of Cambridge, 19 J. J. Thomson Ave.,
Cambridge CB3 0HE, UK}

\author{Xiaojing Lin}
\affiliation{Department of Astronomy,Tsinghua University,Beijing 100084, China }
%\affiliation{Department of Astronomy,
%School of Physics,
%Peking University,
%Beijing 100871, China}
\author{David Brooks}
\affiliation{University College London, Dept. of Physics and Astronomy, Gower Street, London, UK WC1E 6BT}

\author{Solène Chabanier}
\affiliation{Lawrence Berkeley National Laboratory, Berkeley, CA, 94720}

\author{Roger de Belsunce}
\affiliation{Kavli Institute for Cosmology,
University of Cambridge,
Madingley Road, Cambridge CB3 0HA, UK}
\affiliation{Institute of Astronomy, University of Cambridge, Madingley Road, Cambridge CB3 OHA, United Kingdom}

\author{Nathalie Palanque-Delabrouille}
\affiliation{IRFU, CEA, Universit\'e Paris-Saclay, F-91191 Gif-sur-Yvette, France}
%\'e
\author{Gregory Tarle}
\affiliation{Department of Physics, University of Michigan, Ann Arbor, MI 48109, USA}

\author{Zhimin Zhou}
\affiliation{Key Laboratory of Optical Astronomy, National Astronomical Observatories, Chinese Academy of Sciences, Beijing, 100012, China}

\begin{abstract}
%The discovery and characterization of the 
%Damped Ly$\alpha$ systems (DLAs) are crucial in the research of QSO absorption lines. 
We have updated and applied a convolutional neural network (CNN) machine learning model to discover 
and characterize damped Ly$\alpha$ systems (DLAs) based on Dark Energy Spectroscopic Instrument (DESI) mock spectra. 
        We have optimized the training process and constructed a CNN model 
        that yields a DLA classification accuracy above 99$\%$ for spectra which have {\color{black}signal-to-noise (S/N) above 5 per pixel. }Classification accuracy is the rate of correct classifications. 
        This accuracy remains above 97$\%$ for lower signal-to-noise (S/N) $\approx1$ spectra.
        This CNN model provides estimations for redshift and HI column density with standard deviations of {\color{black}0.002 and 0.17 dex for spectra with S/N above 3 per pixel}.
        Also, this DLA finder is able to identify overlapping DLAs and sub-DLAs. 
        Further, the impact of different DLA catalogs on the measurement of Baryon Acoustic Oscillation (BAO) is investigated. 
        {\color{black}The cosmological fitting parameter result for BAO has less than $0.61\%$ difference compared to analysis
        of the mock results with perfect
        knowledge of DLAs. This difference is lower than the statistical error for the first year estimated from the mock spectra: above $1.7\%$.} We also compared the performance of CNN and Gaussian Process (GP) model. %CNN model has more than $5\%$ higher DLA detection rate but GP has a better column density estimate. 
        {\color{black}Our improved CNN model has moderately 14$\%$ higher purity and 7$\%$ higher completeness than an older version of GP code, for S/N $>$ 3. %Performance of the CNN is substantially better for NHI $>$ 22 and  S/N $>$ 3. 
        Both codes provide good DLA redshift estimates, but the GP  produces a better column density estimate by $24\%$ less standard deviation. }A credible DLA catalog for DESI main survey can be provided by combining these two algorithms. 
        %The DLA finder can be mainly used for identification and characterization 
        %of the QSO spectra in the DESI Ly$\alpha$ survey. 

\end{abstract}


\section{Introduction} \label{sec:intro}
%\section{Intro}

The absorption systems in the spectra of the quasi-stellar-objects (QSOs) 
are widely used to probe the properties of the early universe \cite[e.g.,][]{1998ARA&A..36..267R,1986ApJS...61..249W}. 
Using QSO absorption line systems (e.g., Ly$\alpha$ and metal lines), one can probe 
a wide range of scales, including the gas properties inside and around galaxies 
\cite[e.g.,][]{2011MNRAS.418.1796F}. {\color{black}Further, the QSO absorption line 
systems are used to reconstruct the cosmic web on a few tens of Mpc \cite[e.g.,][]{2003ApJ...585...34M, 2016ApJ...833..135C,2014ApJ...795L..12L,2017ApJ...839..131C,2021ApJ...916...20L}, 
and test cosmological models on the cosmological scale of hundreds of Mpc \cite[e.g.,][]{2018MNRAS.473.3019P}. 
Among the absorption systems, %in the QSO spectrum, 
damped Ly$\alpha$ systems (DLAs) are a population of strong absorbers with
integrated neutral hydrogen (HI) column densities $N_{\rm{HI}}$ $>$ $2\times10^{20}$~cm$^{-2}$ \cite[e.g.,][]{2005ARA&A..43..861W}. }
DLAs serve as the dominant reservoirs of atomic hydrogen in the universe and offer a unique opportunity to probe the early universe  \cite[e.g.,][]{1997ApJ...487...73P,2013A&A...556A.141Z}. 
Their absorption can be described using the
Voigt profile which fits the damping wings driven by the natural broadening of the Ly$\alpha$ 
transition\cite[e.g.,][]{2020MNRAS.491.5555L}. 
Recently, DLAs are widely used to probe %can provide us the information about 
the circumgalactic medium (CGM) around %the interstellar medium (ISM) of 
galaxies, especially high-redshift galaxies at $z>2$ \cite[e.g.,][]{2019A&A...627A..32N,1997ApJ...480L..99G}. 
Simulations show that the majority of gas which gives rise to DLAs is associated with galaxies
\cite[e.g.,][]{2014MNRAS.445.2313B,2020arXiv201011254G,2014AGUFM.P51B3932R}. 
A complete understanding of galaxy evolution is based on the analysis for the properties of neutral gas\cite[e.g.,][]{2020MNRAS.495.3014K}. 
Besides, a large sample of QSOs and DLAs can be used for measuring a variety of cross-correlations and auto-correlations, and this could help to fit the Baryon Acoustic Oscillation (BAO) at $z>2$. 
%reveal the Baryon Acoustic Oscillation (BAO) at $z>2$. 

Previously, {\color{black}with the help from visual inspection, \citet{2004PASP..116..622P,2005ApJ...635..123P} searched for DLA candidates in SDSS spectra by running a window along the spectra to identify the DLA trough as a region where the signal-to-noise ratio (S/N) is significantly lower than the characteristic S/N in the vicinity.  
Later, by utilizing a fully automatic procedure based on   
classical statistics, \citet{2009A&A...505.1087N, 2012A&A...547L...1N} identified DLAs in SDSS/DR7 and SDSS-III/BOSS survey. }
Further, with the rapid increase of the spectral data in the era of eBOSS and future surveys, the efficient and accurate detection of DLAs 
from low S/N spectra is becoming a technical challenge. 
{\color{black}An automated technique using Gaussian Process is applied to detect DLAs along QSO sightlines\citep{2017MNRAS.472.1850G,2020MNRAS.496.5436H}. }
Recently, in \citet{2018MNRAS.476.1151P}, a Convolutional Neural Networks (CNN) model was
designed to detect and characterize DLAs in the QSO spectra 
of the SDSS and BOSS survey. 
This algorithm yields a classification accuracy of 99$\%$ on spectra with S/N above five. The classification accuracy is defined as the 
proportion of results with correct predictions.  
%in equation \ref{equ:accuracy}. 
The estimation for column densities and redshifts both have median values consistent with the ground truth, with a scattering of standard deviation of column density of $\sigma(logN_{\rm{HI}}
)$=0.15 and redshift of $\sigma(z)$=0.002, respectively. This CNN model is also applied to SDSS DR16 and a DLA catalog is generated by this algorithm \citet{2021arXiv210709612C}.

DESI (Dark Energy Spectroscopic Instrument) is a stage IV spectroscopic survey project, and it is a 5-year {\color{black}survey for galaxies, QSOs and Milky Way stars}, covering 14,000 deg$^{2}$ \citet{2016arXiv161100036D}. 
The highest redshift coverage of DESI comes from QSOs. 
At higher redshift, DESI will use QSOs as backlights 
to measure clustering in the Ly$\alpha$ forest, 
the series of HI absorption lines in the spectra of distant QSOs. These absorption lines are produced by the Ly$\alpha$ electron transition of the neutral hydrogen \citet{1998MNRAS.301..787L}. 
{\color{black}About 2.4 million QSO spectra are expected to be produced \citet{2020RNAAS...4..179Y}, tracing the 3D distribution of the intergalactic gas at $z\gtrsim2$ with a survey volume of 3 Gpc$^{3}$. }
Comparing with the SDSS survey, it increases the Ly$\alpha$ forest survey volume by an order of magnitude. 
The Ly$\alpha$ forest 
%a collection of HI absorption features in the QSO spectra, 
is now used to provide the BAO measurement at $z\gtrsim2$. 
{\color{black}
\citet{2020ApJ...901..153D} show that the forest with identified DLAs has to be specially treated when BAO analysis is conducted. }DLAs will reduce the flux transmission field for the correlation estimate. 
%in Equation \ref{equ:fluxfield}.
The spectra pixels where a DLA reduces the transmission by more than 20$\%$ should not be used because these pixels will cause a bias in the final BAO measurement. 
%Because the spectra pixels where a DLA reduces the transmission by more than 20$\%$ should not be used. 
That makes a DLA catalog indispensable for precise
and accurate BAO fitting analysis.

Our paper aims to develop a DLA finder for the DESI survey. 
The method adopted is based on the CNN model \citep{2018MNRAS.476.1151P}, 
and we improve the CNN model using DESI mock spectra. 
{\color{black}Different DESI mock 
spectra were chosen to make this algorithm available for a wide range of S/N levels. The minimum S/N of the mock spectra which we used is 0.31.  }
{\color{black}We have 
%optimized the training process and 
updated the framework to TensorFlow2.0. }
Our neural network was trained on a server with two NVIDIA Tesla V100 GPUs. 
After developing the CNN model, we use it to get the DLA catalog for the DESI mock spectra. We have also conducted BAO analysis tests to examine the effects 
of different DLA catalogs with different definitions.

This paper is organized as follows. 
In Section \ref{sec:DLA survey}, we introduce the basic physical features of damping wings and the mock spectra for this paper. 
In Section \ref{sec:train}, the description of the training process is present, including dataset generation, label setting, and training process. 
Model validation is discussed in Section \ref{sec:validation}. 
{\color{black}The comparison of CNN model and Gaussian Process (GP) model \citep{2020MNRAS.496.5436H} on detecting DLAs is discussed in Section \ref{sec:Gaussian process}. }
In Section \ref{sec:BAO}, the DLA catalog is generated. Further, 
the comparison of BAO fitting results using our DLA catalog and mock catalog is quantified. 
All codes related are available at \url{https://github.com/cosmodesi/desi-dlas}

\section{DLA survey} \label{sec:DLA survey}

\subsection{Basic Terminology of DLA}\label{subsec:DLA}

Among all the Ly$\alpha$ absorbers,  DLAs have the highest HI column densities of  $N_{\rm{HI}}\geq 2\times 10^{20} \rm{cm}^{-2}$. 
At lower column densities, we designate absorption systems with $10^{17}\rm{cm}^{-2}\leq N_{\rm{HI}}\leq 2\times 10^{20} \rm{cm}^{-2}$ as Ly$\alpha$ limit systems (LLSs) including sub-DLAs with $10^{19}\rm{cm}^{-2}\leq N_{\rm{HI}}\leq 2\times 10^{20} \rm{cm}^{-2}$ or so-called super Lyman limit systems (SLLSs) \cite[e.g.,][]{2003MNRAS.346.1103P, 2015ApJS..221....2P}. 
At even lower column densities, these systems are called Ly$\alpha$ forest absorbers with $N_{\rm{HI}}\leq 10^{17} \rm{cm}^{-2}$, corresponding to the intergalactic hydrogen “clouds” along the QSO sightline \cite[e.g.,][]{2016ARA&A..54..313M}.
The fundamental difference between DLAs and other Ly$\alpha$ absorbers is that hydrogen is mainly neutral in DLAs, while in all other absorption systems it is ionized \cite[e.g.,][]{2005ARA&A..43..861W}.

{\color{black}DLAs can be fitted by the Voigt profile, 
which is the convolution of Lorentz profile and Gaussian profile \citep{2011piim.book.....D}. }
According to the {\it Uncertainty Principle}, the energy level of an electron has a finite width. 
Therefore when an electron transitions between different energy levels, the corresponding frequency also has a certain range of distribution. 
This broadening is called natural broadening and it can be described by a Lorentz profile. 
Meanwhile, the Doppler effect causes Doppler broadening, which is fitted by a Gaussian profile when %assuming that 
the gas satisfies a Maxwellian velocity distribution. 
The broadening of Ly$\alpha$ absorption lines is mainly caused by natural broadening and Doppler broadening \cite[e.g.,][]{2020MNRAS.491.5555L}. 

At higher column densities, absorbers become optically thicker, and the Voigt profile can be 
characterized by a dark trough and the Lorentz damping wing,  
making it possible to identify individual absorbers from 
even a moderate S/N spectrum.

\subsection{DESI Mock Spectra}\label{subsec:mock}
\subsubsection{Data Sample}
%Since the current sample size of real DESI spectra is not enough to well train the CNN model, 
%we use the mock spectra that are representative of the data quality (e.g., S/N, resolution) of DESI. 
The DESI mock spectra are representative of the data quality  (e.g., S/N, resolution) of DESI real data. 
The location of DLAs and their column density are %clearly 
known in mock spectra, so the mock spectra can be used to 
%well 
train the CNN model and evaluate its performance. 
We choose four mock spectral database (mocks) 
from {\color{black}LyaCoLoRe mocks} %LondonMock v9.0 
generated by the DESI Ly$\alpha$ Forest Working Group \citep{2020JCAP...03..068F}. 
To add DLAs in the mock spectra, {\color{black}the location of DLAs can be derived by a Gaussian field which is used to compute the density and velocities in the spectra \citep{2012JCAP...07..028F}. }
Then a column density is allocated to each DLA  {\color{black}following the observed column density distribution from \texttt{pyigm} \footnote{Publicly available at \url{https://github.com/pyigm/pyigm.}}}, and the absorption profile is calculated using a Voigt template. 
{\color{black}After these steps are done in the LyaCoLoRe mock production stage \citep{2020JCAP...03..068F}, DLAs can be inserted into the final synthetic spectra.}
%After these steps, DLAs can be inserted into the final spectra . 
Table \ref{tab:mock} lists the information of the four mock database we used. 
\begin{deluxetable}{ccccc}
\tablecaption{Information of four DESI Mock \label{tab:mock}}
\tablehead{
\colhead{Name\tablenotemark{a}} & \colhead{Exposure Time } &\colhead{DLAs} & \colhead{Metals} & \colhead{BALs}\\
\colhead{} & \colhead{(1000s)} & \colhead{} & \colhead{} & \colhead{}  
}
\startdata
desi-0.2-100 & 100 & Y & Y & N \\
desi-0.2-4 & 4 & Y & Y & N \\
desi-0.2-1 & 1 & Y & Y & N \\
desiY1-0.2-DLA & multiple & Y & N & N \\
\enddata
\tablenotetext{a}{These Mocks are named: name-ver-nexp, where name is a short name to differentiate between different sets of runs, ver determines what version of systematics have been used, and nexp determines the number of exposures. }
\end{deluxetable}

The first three mocks have the 
same QSO catalog with different exposure times. 
The redshift of QSOs in the mock catalog is  {\color{black}from 1.8 to 3.8}. {\color{black}The continuum of these QSOs are generated using the publicly available package {\it{simqso}} 
as implemented in {\it{desisim}} code  \footnote{https://github.com/desihub/desisim}. The basic procedure is to generate an unabsorbed continuum for each QSO by adding a set of emission lines on top of a broken power law continuum model.  The {\it{simqso}} is based on \cite{2013ApJ...768..105M}, however for these DESI mocks the emission lines had been tuned to provide a similar mean continuum, in the Lyman-$\alpha$ and Lyman-$\beta$ forest region, to that observed in eBOSS DR16 \citep{2020ApJ...901..153D}. A wider description of the {\it{desisim}} implementation to generate the continuum, as well as the full synthetic spectra production itself, including the DLA insertion, will be presented in detail in \citet{Gonzalez-Morales2021} %Gonzalez-Morales et al. (2021, in preparation) %\cite{gonzalez-morales2021}
; it is worth noticing that eBOSS DR16 used a quite similar mock set. }
%The continuum of these QSOs are generated from the QSOs in SDSS's Baryon Oscillation Spectroscopic Survey (BOSS), using Principal Component Analysis (PCA). 
%Details for this procedure are shown in desisim \url{https://github.com/desihub/desisim}. 
There are many different versions of mock spectra in DESI, we choose some versions that the mock spectra are inserted with DLAs to do the training. 
%the mock spectra inserted with DLAs are chosen for our training,
{\color{black}
For the spectra we used, the marked number '0.2' \footnote{0.0 no extra systems.  0.2 with DLAs. } means that these spectra are inserted with DLAs. 
%but exposure time is limited.   %0.3 with BALs. 
%The number `0.2' stands for 
%these spectra were inserted with DLAs, 
The last marked number (such as '100', '4', '1') stands for the exposure time these mock spectra have. 
The mock spectra 'desi-0.2-100' have the S/N level equal to the spectra with exposure time $10^5$s. 
This is the most noise free sample in our paper. }
{\color{black}For 'desi-0.2-1' and 'desi-0.2-4' mocks, simulated QSO spectra have a fixed S/N level similar to DESI spectra observed one or four DESI effective times,  1000s and 4000s respectively.}
%The mock spectra 'desi-0.2-1' and 'desi-0.2-4' have the S/N level similar to DESI spectra in the first year and four years combined respectively. 

{\color{black}The last mock, DesiY1-0.2-DLA is more realistic of what we would expect from the DESI first year observations, since those were constructed using a survey simulation to determine what region of the DESI footprint \citep{2019AJ....157..168D} would be covered during DESI's first year, given a random realization of observing conditions. Such survey simulation also includes a similar target selection criteria as the main DESI Survey and a simplified fiber assign procedure to reflect that  high redshift QSO can be observed up to four times (of ~1000s each), as opposed to most target that are observed only once,  depending on what other targets are available to be observed and whether we know the QSO redshift with high significance. This procedure results in a mock spectra sample of low redshift and high redshift QSOs which have a distribution of exposure times ranging from 1000s to 4000s, i.e it has a more realistic S/N distribution. These simulations were done using several pieces of desicode \footnote{https://github.com/desihub/} and were presented in \citet{Herrera-Alcantar2020}.}
%The last mock, DesiY1-0.2-DLA is more realistic as the DESI survey footprint of the first year \citep{2019AJ....157..168D}, densities of low redshift and high redshift QSOs, and the exposure times distribution are adopted. 

{\color{black}The BAL (broad absorption lines) features have similar profile to DLAs. We do not simulate BAL features in all the mock we used. This is to avoid their confusion on the training and reducing
false-positive predictions of the model.  }
%, whose profile is similar to DLAs, are excluded from all the mock spectra {\color{black}we used}, avoiding their confusion on the training and reducing
%false-positive predictions of the model. 
%However, this implies the CNN model will be prone to misidentify BALs as DLAs.

\subsubsection{Data Structure of DESI Spectra}

The DESI spectrometer uses three cameras to measure the flux in different wavelength channels. 
Every spectrum in the same camera shares the same wavelength array. 
The three channels are given in Table \ref{tab:rebin}.

\begin{figure}

\plotone{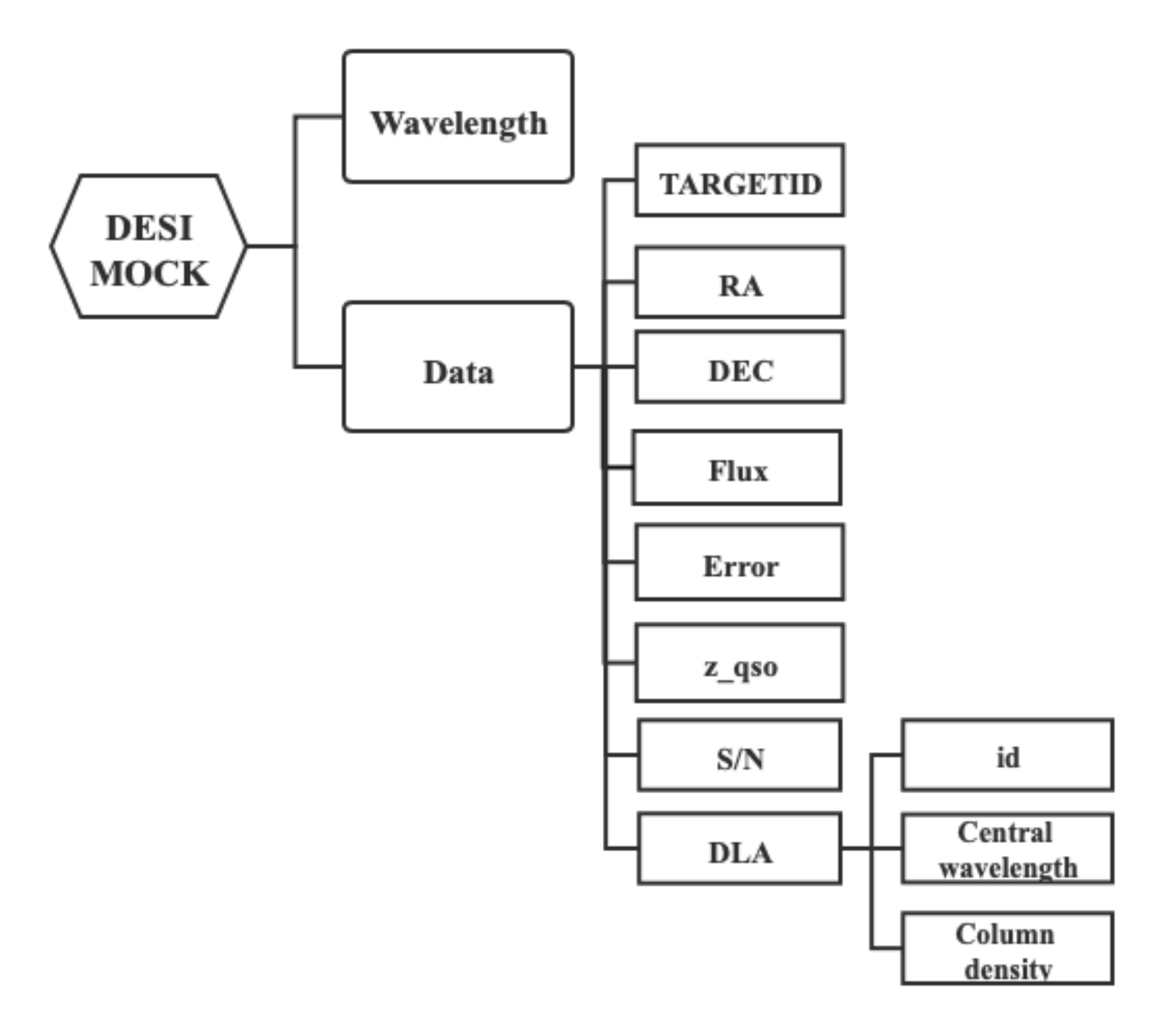}
\caption{In the data structure of the DESIMOCK class, the wavelength array and the information of every spectrum are stored separately, including TARGETID, RA, DEC, etc. \label{fig:DESIMOCK}}
\end{figure}

We used a class named DesiMock to store the various information of every spectrum. 
As shown in Figure \ref{fig:DESIMOCK}, using the spectrum TARGETID as the index, each spectrum contains flux, error array, celestial coordinates, QSO redshift, S/N, and 
the DLA information, which contains the DLA ID, DLA central wavelength, and neutral hydrogen column density ($N_{\rm{HI}}$). 
We read the spectral data from this DesiMock class. 

\subsubsection{S/N Definition}

\begin{figure*}
\plotone{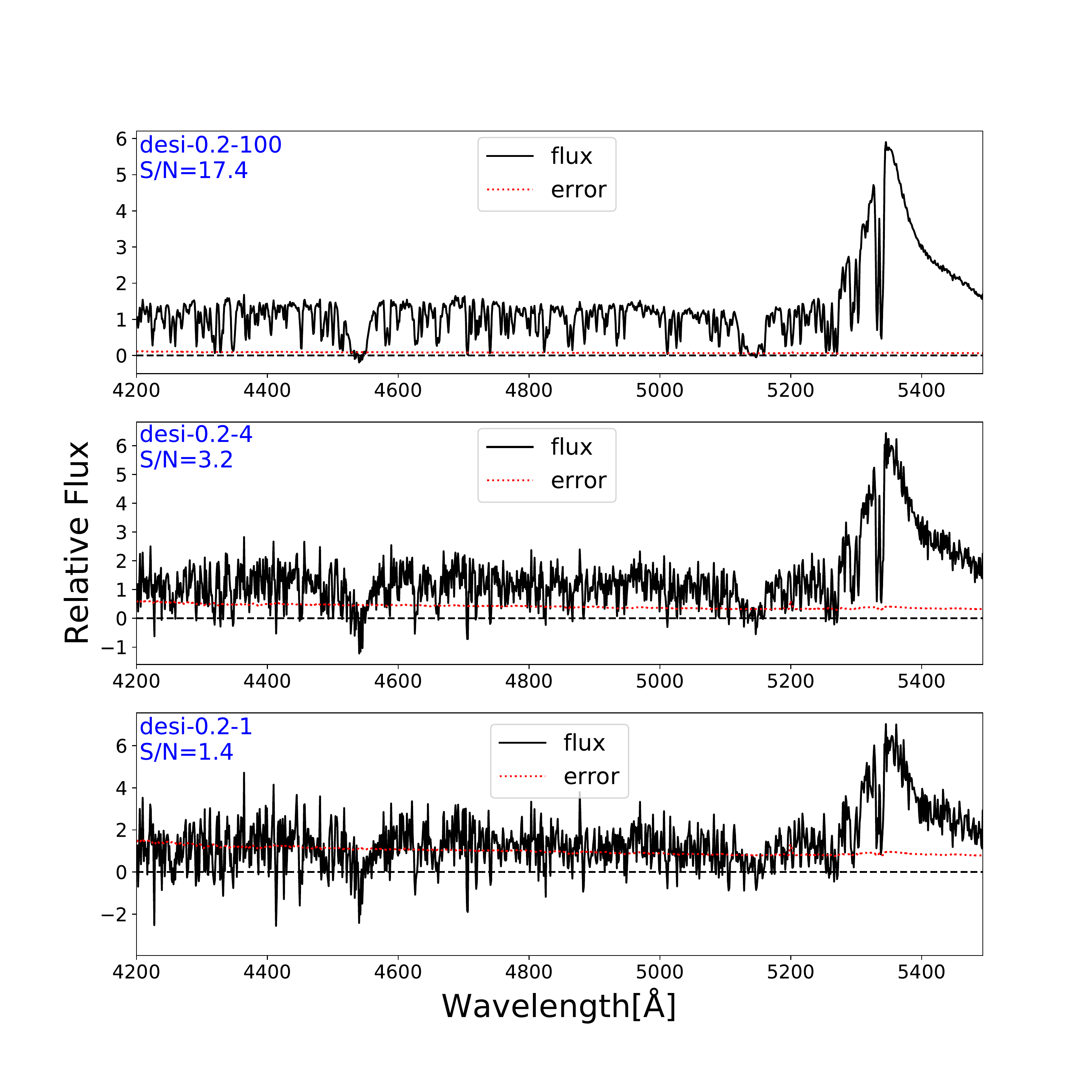}
\caption{The three mock spectra of a QSO at $z_{em}\approx3.395$
which exhibits two DLAs at $z_{\rm{abs}}\approx 2.733,3.330$ with $N_{\rm{HI}} \approx 10^{20. 69} \rm{cm^{-2}},10^{20. 23} \rm{cm^{-2}}$, which can be seen on the graph at the wavelength $\lambda_{\rm{abs}}\approx4538 \rm{\AA},5142 \rm{\AA}$ respectively. The signal-to-noise (S/N) is shown at the upper left of each panel. The black line is the flux and the red dashed line is error.  \label{fig:3 sightlines}}
\end{figure*}

Figure \ref{fig:3 sightlines} shows three spectra with different S/N of the same QSO 
(Mock spectrum ID: 170257611) with emission redshift $z_{em}\approx3.395$. 
The S/N of the spectra in Figure \ref{fig:3 sightlines} is estimated from the median flux of the data to the error array. 
Note the QSO rest-frame 1420\rm{\AA} to 1480\rm{\AA} do not have strong line features, and 
thus, we define the S/N of each mock spectrum as follows:  
\begin{equation}
\rm{S/N}_{\rm{sightline}}= \rm{median}\left(\frac{{\rm flux(\lambda)}}{{\rm error(\lambda)}}\right), \ \lambda=1420-1480\rm{\AA}
\end{equation}
{\color{black}The S/N definition is for per pixel. The following S/N values in this paper are all per pixel. }
Note that the S/N of the Ly $\alpha$ forest is much lower than the S/N defined here. For example, for a spectrum with S/N $=$ 3, the typical S/N in the forest region could be 
lower than unity. 

\begin{figure*}
\plotone{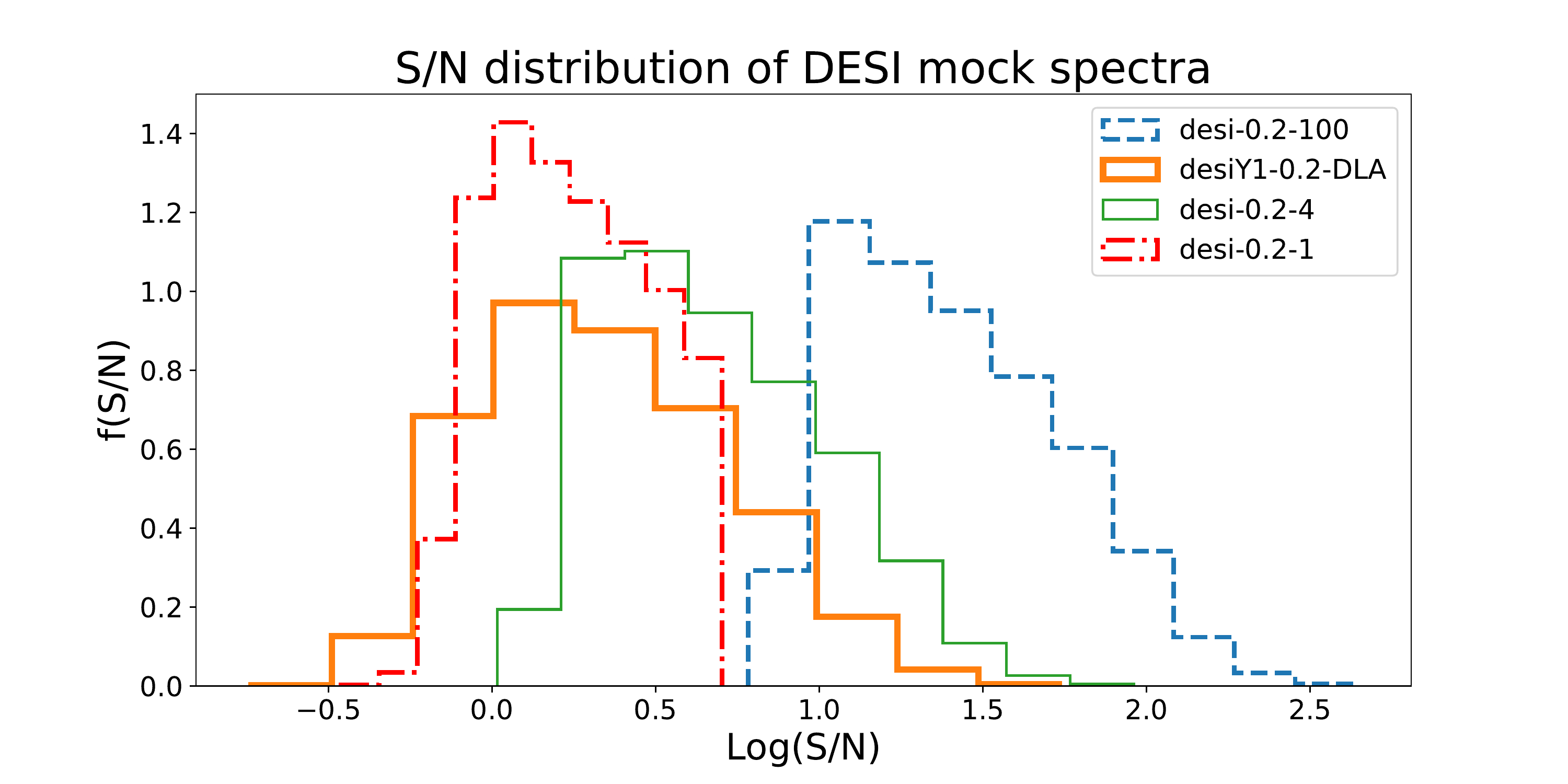}
\caption{S/N distribution of four mocks. The three different mocks: desi-0.2-100, desi-0.2-4, desi-0.2-1  are used to train the CNN model at different S/N. It can be seen that the mock desiY1-0.2-DLA has more spectra with S/N$<$1. \label{fig:S/N distribution}}
\end{figure*}

Figure~\ref{fig:S/N distribution} displays the S/N distribution of the four mock datasets. 
The spectra used for training and prediction should have similar S/N ratios. 
Therefore, we used the first three mocks to train models
for each mock respectively, 
as described in Section \ref{sec:train}. The last mock desiY1-0.2-DLA is used to validate the three models, 
as described in Section \ref{sec:validation}. 

\section{Training} \label{sec:train}

\subsection{Preprocessing} \label{subsec:preprocessing}

\subsubsection{Rebin}

Spectral rebin consists of changing the size of the spectral bins of each spectra 
depending on the width of the line \citep{2020MNRAS.499.3992J}. 
The dispersion $\Delta\lambda$ of the DESI mock spectra is 
$\approx 0.8 \rm{\AA}$ per pixel, and hence the resolution is not a constant along the spectrum  ($R=\frac{\lambda}{\Delta\lambda}$). 
The instruments for DESI cover
wavelength range from 360nm to 980nm with resolution R=2000–5500 depending on wavelength\citep{2016arXiv161100037D}. 
This means that in the DESI spectra, 
{\color{black}the number of pixels that span a DLA feature with a given $N_{\rm{HI}}$ is proportional to their redshift. }
A DLA at higher redshift has more pixel numbers than a DLA with same column density but at lower redshift. 
This will affect the model estimation of the HI column density. 
%but note that the pixel number could affect the model estimation of the HI column density.
To correct this effect, we have to make sure the pixel size is a constant in 
the velocity-space. Thus, we set: 

\begin{equation}
    \frac{\Delta\lambda}{\lambda}=ln(1+\frac{\Delta v}{c})
\end{equation}
where the $\Delta \lambda$ represents the dispersion per pixel, and $\Delta v$ represents the median pixel size in velocity. 
Then, we interpolate the original grid to the rebinned new grid with the pixel size equal to $\Delta v/c$. 
As seen in Table~\ref{tab:rebin}, we set the pixel size as the median velocity value in each channel.

\begin{deluxetable}{cccc}
\tablecaption{Value of resolution for three channels \label{tab:rebin}}
\tablehead{
\colhead{Channel} & \colhead{Blue Channel} &\colhead{Red Channel} & \colhead{Z Channel} 
}
\startdata
Wavelength($\rm \AA$) & 3570-5950 & 5625-7740 & 7435-9833 \\
$\Delta$ v (km/s) & 63.0  & 44.9 & 34.7 \\
\enddata
\end{deluxetable}

%v (km/s) & 62.996  & 44.859 & 34.720 \\

\subsubsection{Generating Datasets for Training and Validation} \label{subsec:Generating Dataset}

Previous DLA surveys \citep{2009A&A...505.1087N} 
firstly estimate the QSO continua and then normalize the flux. 
Nevertheless, \citet{2018MNRAS.476.1151P} claimed that CNN learned to account for the continuum during the training. The CNN model could potentially detect DLAs without modeling the QSO continuum. Thus, we only use the median flux in the interval of 
1420\rm{\AA} to 1480\rm{\AA} 
to do the normalization.
Then, we construct the appropriate flux dataset for training and validation. 
The sightlines are processed in the following paragraphs.  
Similar treatment can be found in \citet{2018MNRAS.476.1151P}.

\begin{figure*}
\plotone{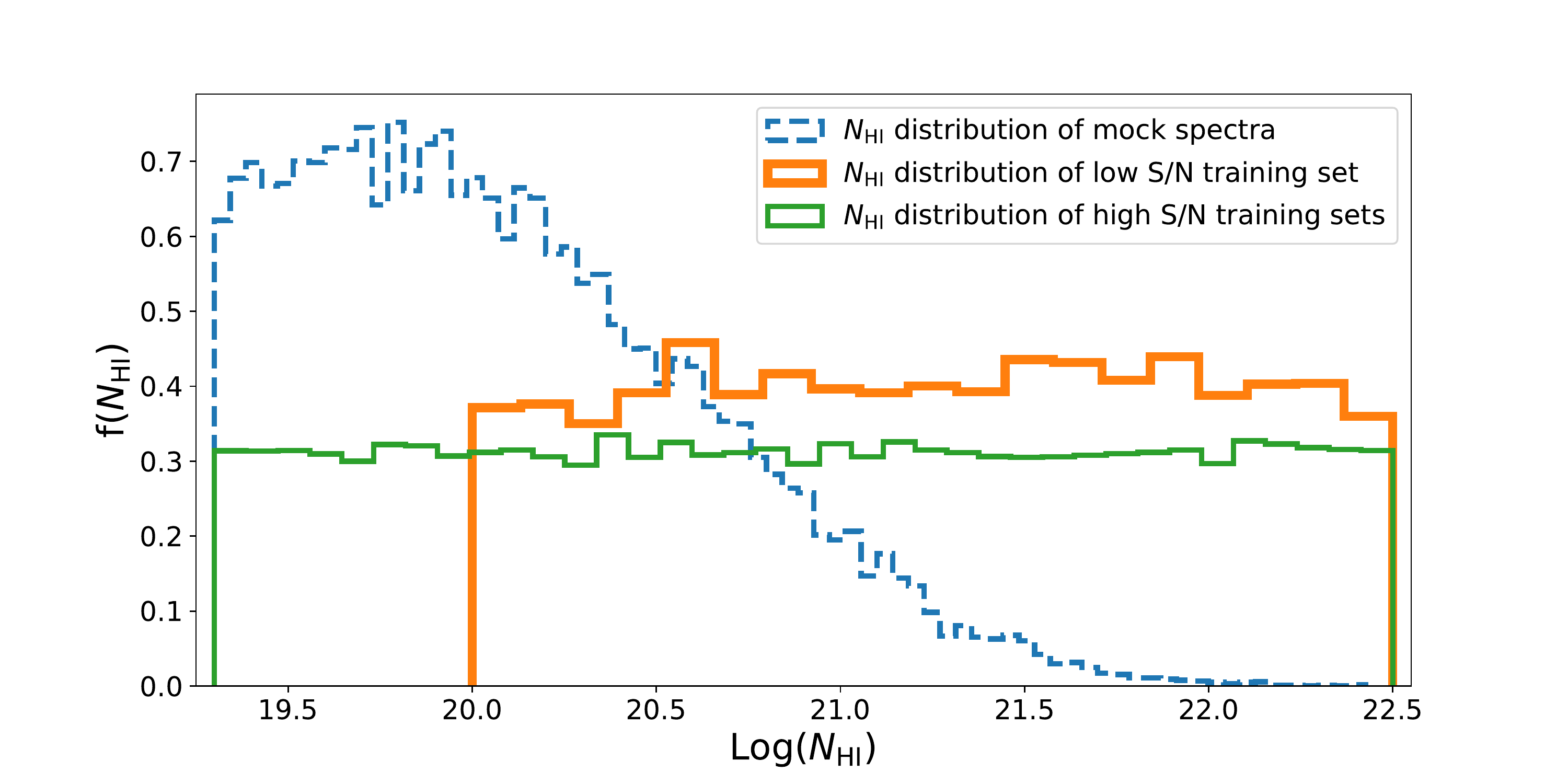}
\caption{$N_{\rm{HI}}$ distribution of mock spectra and training sets. There are more DLAs with low column density in the mock data which follows the empirically measured
distribution. The distribution of column density in training data is uniformed to reduce bias of the model
to lower $N_{\rm HI}$ values. \label{fig:NHI distribution}}
\end{figure*}

\begin{itemize}
\item We only use a fixed range of the sightline ranging from 900~\rm{\AA} to 1346~\rm{\AA} in the QSO rest frame. 
The lower bound ensures that intervening optically thick HI gas below 900\rm{\AA} does not affect the identification of DLAs, and the upper bound ensures that DLAs on or near the QSO Ly$\alpha$ emission can be recovered. {\color{black}The purpose of choosing 1346 as an upper limit is to avoid missing some high column density associated DLAs which can blocks the Broad Line Region emission from the QSOs \citet{2013A&A...558A.111F} (for example, DLAs with $N_{\rm{HI}}>10^{22}$ and less than 1500$km s^{-1}$ from the QSOs redshift). }\textcolor{black}{Among more than 40000 DLA candidates detected by CNN in desiY1-0.2-DLA mock spectra, only 79 DLAs are located above the rest-frame 1216\AA, and they can be excluded using the redshift cut. }

\item 
Each spectrum contains more than 2000 pixels. 
Inputting the whole sightline directly into the model leads to difficulties in 
discriminating multiple DLAs. 
Thus, we input a sliding window of a fixed pixel regions centered on each pixel into the model.
The choice of window size and the hyperparameter selection process are discussed in detail in Section \ref{subsubsec:Hyperparameter Search}.

\item {\color{black}Since there are far more regions without DLAs than regions with DLAs, our training sets maintain a 50/50 balance between training on positive and negative regions. This means the training datasets have half regions with DLAs and half regions without DLAs. This can help to train the CNN model on both positive and negative samples. }

\item Some regions of spectra are not included in the datasets. 
In the training set, we exclude the fixed pixels regions centered on DLA boundary and Ly$\beta$ absorption regions. {\color{black}60 pixels on DLA boundary are avoided in the training set. We mask 15$\AA$ region around the Ly$\beta$ absorption. }When we label the dataset, the classification value changes abruptly from 1 to 0 on the DLA boundary, which confuse the model. 
The Ly$\beta$ absorption lines corresponding to DLAs may be incorrectly detected as DLAs coming from the lower redshift.

\item Flatten the column density distribution of SLLSs and DLAs. 
The dashed line of Figure \ref{fig:NHI distribution} shows the $N_{\rm{HI}}$ distribution of the 
mixed mock spectra. The mixed spectral database is combined with different DESI mock catalogs including desi-0.2-100, desi-0.2-4 and desi-0.2-1. There are far more low $N_{\rm{HI}}$ DLAs than high $N_{\rm{HI}}$ DLAs which would induce bias towards to lower value when estimating column density in the algorithm. 
Thus, we manually inserted DLAs and super Lyman limit systems (SLLSs) into sightlines without high column density systems (HCDs), following the method described in Section 4.2 of \citet{2018MNRAS.476.1151P}. 
The redshift distribution for the inserted DLAs is uniformed to avoid bias in training.  
{\color{black}The final $N_{\rm{HI}}$ distribution of our training sets is uniform with log $N_{\rm{HI}}$ ranging from 19.3 to 22.5 for multiple exposure time mocks (desi-0.2-100 and desi-0.2-4) and from 20.0 to 22.5 for single exposure time mock  (desi-0.2-1), shown in the solid line of Figure \ref{fig:NHI distribution}. If we do not make the log $N_{\rm{HI}}$ uniformed, the CNN model will perform a bias for $N_{\rm{HI}}$ estimate. The  $N_{\rm{HI}}$ distribution for training is uniformed to avoid the bias. }

\end{itemize}

Following these procedures, we generate DLA training samples. 

In Table \ref{tab:trainingset}, we list the number of sightlines containing different absorbers for the training sets. 
With the same distribution as training sets, we generated the validation sets for Section \ref{sec:validation}, see Table \ref{tab:validationset}.

\begin{deluxetable*}{cccccc}
\tablecaption{Information of Training Sets \label{tab:trainingset}}
\tablehead{
\colhead{S/N Level}  &\colhead{DLAS}  & \colhead{high $N_{\rm{HI}}$ DLAs\tablenotemark{a}} &
\colhead{total number of sightlines} & \colhead{total number of absorbers} & \colhead{mock name}
}
\startdata
S/N 1-3 & 50893 & 34128 & 45748 & 57970 & desi-0.2-1 \\
S/N 3-6 & 48068 & 17099 & 100000 & 127347 & desi-0.2-4   \\
S/N 6-  & 58453 & 39453 & 65369 & 85436 & desi-0.2-100 \\
\enddata
\tablenotetext{a}{Here we point out high $N_{\rm{HI}}$ DLAs with log$N_{\rm{HI}}>21.0$.}
\end{deluxetable*}

\begin{deluxetable*}{cccccc}
\tablecaption{Information of Validation Sets \label{tab:validationset}}
\tablehead{
\colhead{S/N Level} &\colhead{DLAS} & \colhead{high $N_{\rm{HI}}$ DLAs\tablenotemark{a}} &
\colhead{total number of sightlines} & \colhead{total number of absorbers} & \colhead{mock name}
}
\startdata
S/N 1-3  & 5366 & 3657 & 4938 & 6040 & desi-0.2-1 \\
S/N 3-6  & 20365 & 7267 & 43224 & 55166 & desi-0.2-4 \\
S/N 6-  & 6539 & 1360 & 47424 & 19917 & desi-0.2-100 \\
\enddata
\tablenotetext{a}{Here we point out high $N_{\rm{HI}}$ DLAs with log$N_{\rm{HI}}> 21.0$.}
\end{deluxetable*}

\subsubsection{Improvement on Low S/N spectra}

\begin{figure*}
\plotone{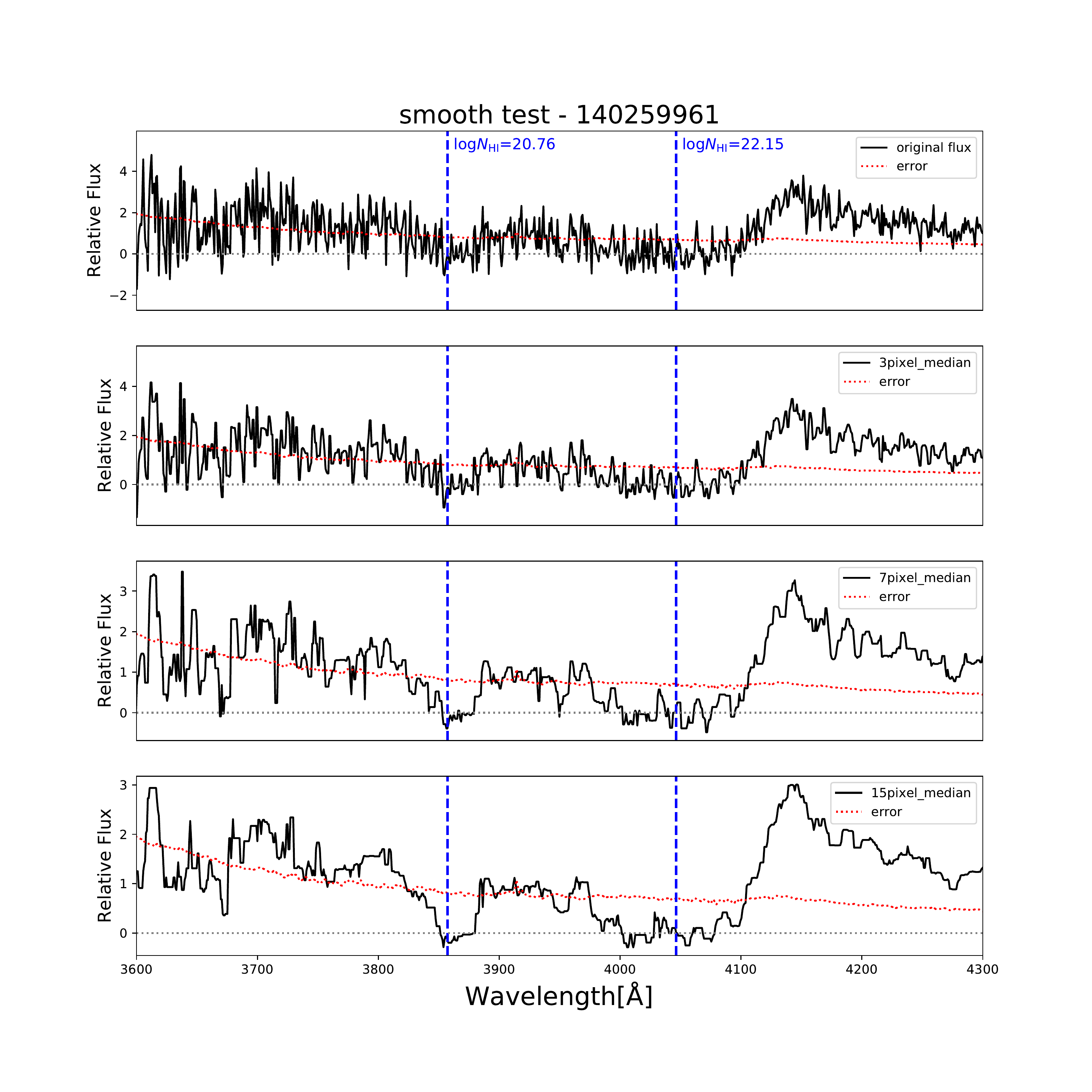}
\caption{Smoothing result for the spectra. The blue dashed line stands for the center of DLAs. The red dashed line is the error for the spectra. The upper panel is the original flux. The lower three panels show the flux after median smoothing for 3 pixels, 7 pixels, and 15 pixels.  \label{fig:Smoothing result}}
\end{figure*}

More than $70\%$ of the mock spectra have S/N$<$3.
The classification accuracy for these low S/N spectra is only $93\%$ using the initial model. 
To improve the accuracy, we used the median smoothing method to optimize the preprocessing. 
Smoothing spectra reduces the resolution but it improves the S/N level. 
Accordingly, for dealing with low S/N spectra, 
we set the training data as a two dimensional array  ($600\times4$). 
The first row is the original flux.
The other three rows are the median smoothing results for 3 pixels, 7 pixels, and 15 pixels. 
The CNN model is adjusted according to this new training data. 
The smoothing process and result are shown in Figure \ref{fig:Smoothing result}. 

\graphicspath{ {. /figures/} }

\subsection{CNN Structure and Model Training} \label{subsec:CNN architecture}
\begin{figure*}
\plotone{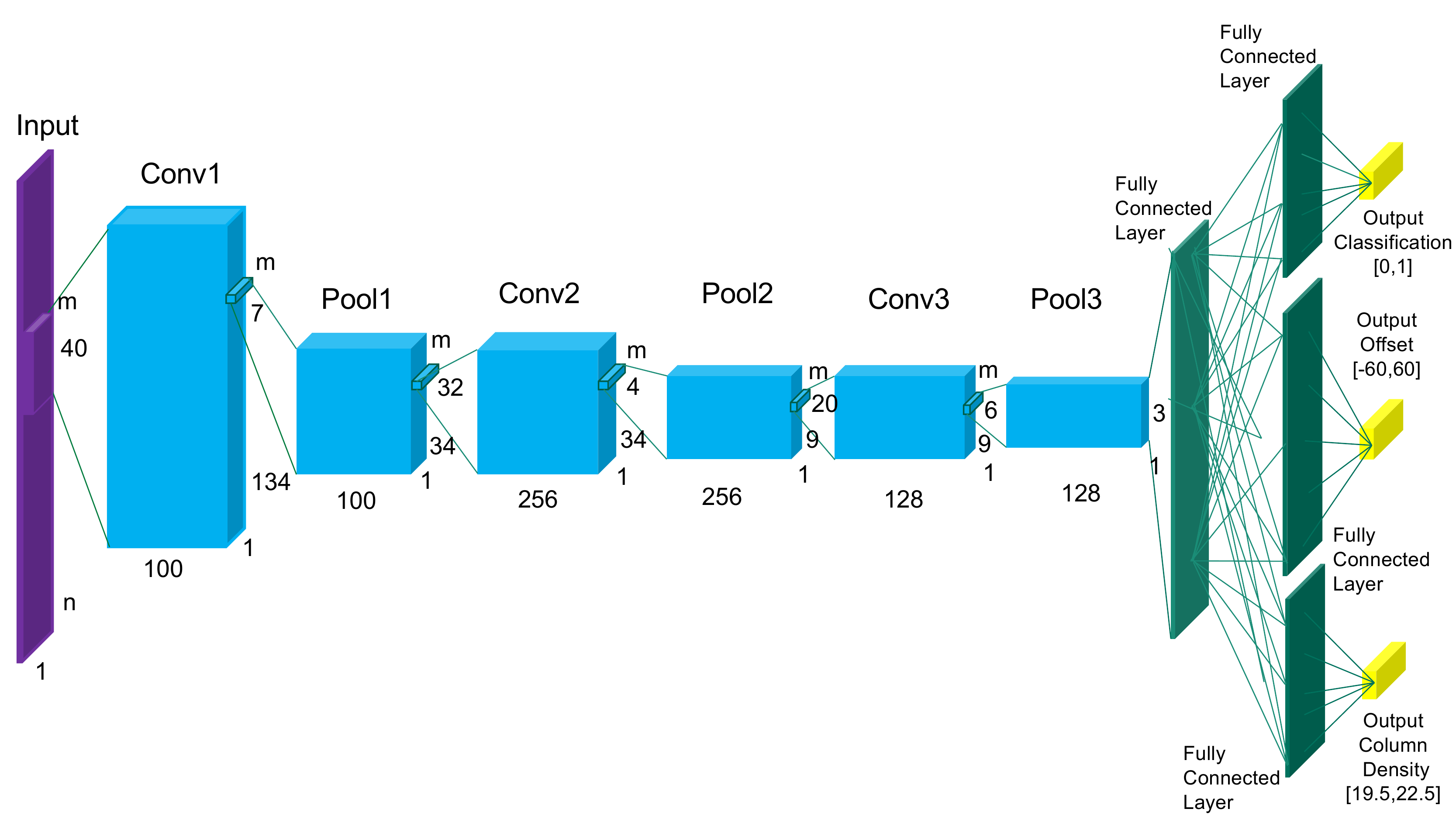}
\caption{The standard CNN architecture we used, casting DLAs as a 1D image problem. {\color{black}There are three convolutional layers, three pooling layers, one fully connected layers, and three sub-fully connected layers. }We have reset two parameters variable. One is $n$, it stands for how many pixels one window contains, it could be 400 or 600. The other one is $m$, it means the dimensions of the input data, the value of it is 1 or 4 (median smoothing for low S/N spectra). The three sub-fully connected layers are correspond to three labels: classification, offset and column density. \label{fig:CNN structure}}
\end{figure*}

We followed the standard CNN architecture constructed in \citet{2018MNRAS.476.1151P}. 
As shown in Figure \ref{fig:CNN structure}, this model has three convolutional layers, each with a max pooling layer, 
following a fully connected layer and the last layer containing 
three separate fully connected sub-layers.

We trained our model 
for $10^6$ iterations each time. 
During the training process, 
we recorded the training accuracy every 200 steps and testing accuracy every 5000 steps. 
The classification accuracy 
improves more than 90$\%$ at the first $10^5$ steps. 
But it achieves the best accuracy of 99$\%$ (for S/N$>5$ spectra, spectra with different S/N levels have different best accuracy)
at about $8\times10^5$ steps.
Then, the accuracy improves less than $0.001\%$ for the rest of the steps.
Thus, $10^6$ iterations are enough for this training. 
The final classification accuracy for different S/N spectra are shown in Table \ref{tab:accuracy}. The training accuracy is the classification accuracy during training, and the definition of testing accuracy is similar. After the smoothing adjustments, the classification accuracy rises from 93$\%$ to 97$\%$ for spectra with S/N$<$3. 

\begin{deluxetable}{ccccccc}
\tablecaption{Training and Testing Accuracy \label{tab:accuracy}}
\tablehead{
\colhead{S/N Level} & \colhead{Training Accuracy } &\colhead{Testing Accuracy} 
}
\startdata
S/N $>$6 & 99$\%$ & 99$\%$ \\
S/N 3-6 & 98$\%$ &97$\%$  \\
S/N 1-3 & 94$\%$ & 93$\%$ \\
S/N 1-3(smooth) & 97$\%$ & 97$\%$ \\
\enddata
\end{deluxetable}

The definition of classification accuracy is based on the confusion matrix Table \ref{tab:confusion matrix}. The label in this confusion matrix is the 'pred' label as described in Section \ref{subsec:CNN architecture}. The classification accuracy is defined as follows:

\begin{deluxetable}{ccccc}
\tablecaption{Confusion Matrix \label{tab:confusion matrix}}
\tablehead{
\colhead{Label} & \colhead{GroundTruth 0 } &\colhead{GroundTruth 1} 
}
\startdata
Prediction 0 & True Negative (TN) & False Negative (FN)  \\
Prediction 1 & False Positive (FP) & True Positive (TP) \\
\enddata
\end{deluxetable}

\begin{equation}
Accuracy = \frac{TP+TN}{TP+FP+TN+FN}
\label{equ:accuracy}
\end{equation}

\begin{figure*}
\plotone{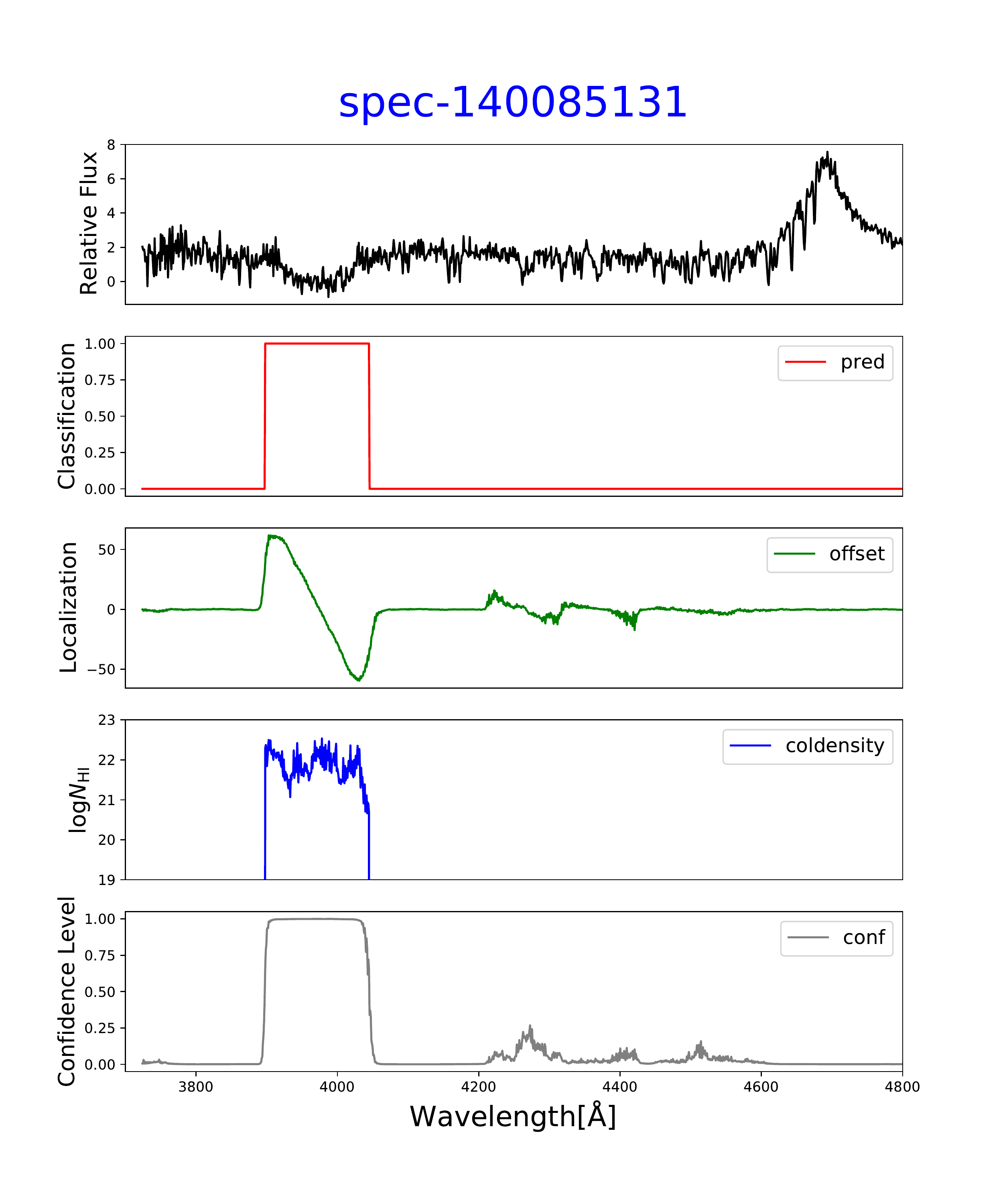}
\caption{Four outputs for every window: 'pred', 'offset', 'coldensity' and 'conf' .  The red line is 'pred', the value for 120 pixels near the center of a DLA is 1 and for other pixels is 0. The green line is 'offset', it shows the distance between every pixel and the center pixel of a DLA. The value of offset is close to 0 for the pixles without a DLA. The blue line is the column density. The last grey line is the confidence level. This is an example with high confidence level, the confidence value for pixels with a DLA is above 0.9.    \label{fig:analyse}}
\end{figure*}

This model yields four outputs for every window of the spectrum, three labels from the full connected layers as shown in Figure \ref{fig:CNN structure} and the confidence level:

\begin{enumerate}
\item {\color{black}'Prediction', labeled as 'pred' in the code. }
This is the classification of the DLA. 
The value of this label is 0.0 or 1.0 . 
1.0 stands for that the model detects a DLA in this window and 0.0 means no detection. 
\item {\color{black}'Offset', labeled as 'offset' in the code. }
This is similar to the generated label. 
It stands for the distance between DLA center and the spectra window center. 
The value of this output is in the range [-60,+60]. 
If there is no DLA in this window, this label will be 0. 
\item {\color{black}'Column Density', labeled as 'coldensity' in the code. }This is the predicted column density of DLAs. 
This label is 0 if the model predicts that no 
DLA lies in this window. 
\item {\color{black}'Confidence Level', labeled as 'conf' in the code. }This output is not from the fully-connected sub-layers. It is not necessary for the prediction but will be helpful for the analysis. It is the confidence level for each prediction. 
It represents the possibility that one DLA is located in this window\citep{2018MNRAS.476.1151P}. 
There are similar concepts in \citet{2012A&A...547L...1N,2012A&A...540A..63N}. 
{\color{black}The value of label `pred' (0 or 1) is determined by this 'conf' label. $C_{\rm min}$ is defined as the minimum of confidence level of a DLA. 
In the original set, 
%$C_{\rm min}$
if label `conf' is above $C_{\rm min}$, then the label `pred' will be 1. 
The critical value of $C_{\rm min}$ is set as 0.5.  }
Please note that this critical value is adjustable. 
By changing this critical value, we can optimize the model prediction,  especially 
for low S/N spectra. 
\end{enumerate}
These four different output labels are also shown in  Figure \ref{fig:analyse}.

\subsection{Hyperparameter Search} \label{subsubsec:Hyperparameter Search}

There are 26 parameters for this CNN model which determines the size of each layer. We have conducted the 
hyperparameter search for these parameters. Since we optimize the hyperparameters for the DESI mock spectra, 
our best combination of the hyperparameters is different 
from that of \citet{2018MNRAS.476.1151P} which is optimized for SDSS spectra.  

A normal important parameter in this algorithm is the input size of the spectral window. 
In \citet{2018MNRAS.476.1151P}, each QSO spectrum was cut into windows with the size of 400 pixels, 
and the classification accuracy depends on the size of the window. 
We have measured the classification accuracy by changing the window  
size from 300 pixels to 700 pixels, with the step of 100 pixels. 
We find that a window size of 600 pixels, combined with the hyperparameters determined above, 
giving the best accuracy.

\section{Validation} \label{sec:validation}

\subsection{Purity and completeness} \label{subsec:Classification Accuracy}
Beside the classification accuracy, the purity and completeness are also important results for the CNN model. 
Purity and completeness are both for DLAs because TN samples are excluded. 
The confusion matrix defination is similar to Table \ref{tab:confusion matrix} but without TN samples. 
GroundTruth stands for the label of DLAs in mock spectra and the prediction is the DLA label from our CNN. 
A DLA in mock spectra will have the GroundTruth value as 1. 
Similarly, the prediction of 1 means a DLA 
is detected by the CNN model. 
If our CNN missed one DLA, then a FN sample is produced (GroundTruth as 1 but prediction as 0). 
After the prediction, two DLA catalogs are generated. One is produced by the CNN model and another one is the mock DLA catalog.  
For each DLA in the mock catalog, we search DLAs in the same sightline among the predicted DLA catalog and then compare the distance between the center of these two DLAs. 
{\color{black}If the distance is less than 10 $\rm{\AA}$, this is a true positive prediction and these two DLAs will be both marked as TP. 
We choose 10 $\rm{\AA}$ as the critical value to make the redshift estimate for TP DLAs has less than 0.008. This difference in redshift estimate is acceptable. This critical value can make the CNN model provide accurate redshift estimate ($<0.008$) and high classification accuracy ($> 97\%$) when comparing the result to mock spectra. 
If the distance is above 10 $\rm{\AA}$, this is a false negative prediction and the DLA in mock catalog will be marked as FN. }
After that, DLAs in predicted catalog without a TP label will be considered as a FP sample. 
The purity and completeness are defined as
%To make this concept more clear,
%we will firstly define the confusion matrix as shown in Table  \ref{tab:confusion matrix}. 

%0 means there is no DLA in this spectra. 

\begin{equation}
Purity=\frac{TP}{TP+FP}
\end{equation}
\begin{equation}
Completeness=\frac{TP}{TP+FN}
\end{equation}

%The accuracy is for the sightlines because it includes the TN samples while the purity and completeness are both for DLAs because TN samples are not taken into account. 
By changing the critical value of $C_{\rm min}$, we obtain different DLA catalogs. 
The default critical value is 0.5 as introduced in Section \ref{subsec:CNN architecture}. 
Changing this critical value changes the FP rate and FN rate, and yields different DLA catalogs. 
The higher critical values improve the purity but reduce the completeness. To balance the purity and completeness, this value is still set as 0.5 in our model. 

\begin{figure*}
\plotone{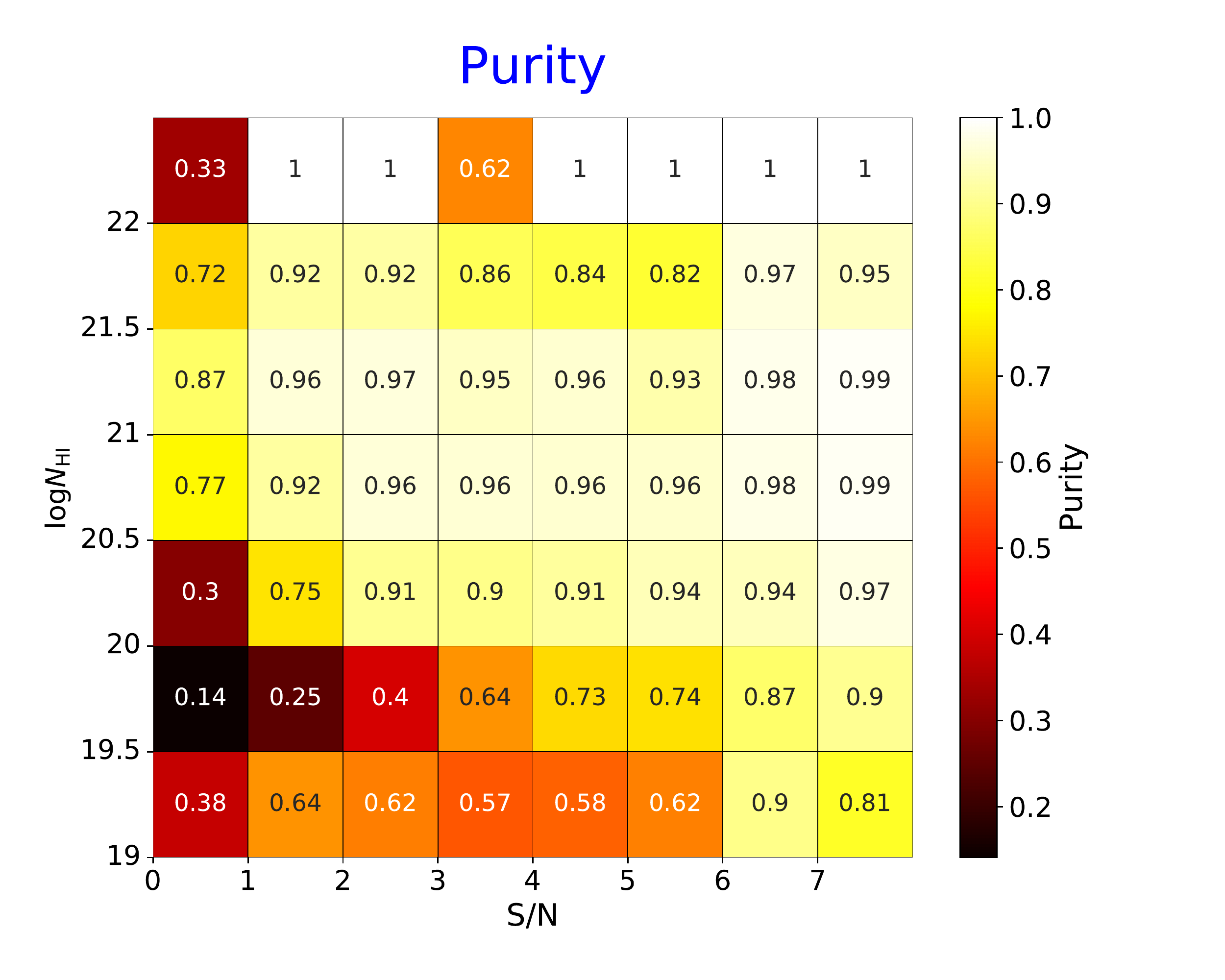}
\caption{Purity results for different S/N levels and column densities using desi-Y1 mock spectra. This is the result choosing the critical value of $C_{\rm min}$ as 0.5. %The lower left bins are blank because we do not train our model to detect sub-DLAs in low S/N (S/N$<3$) spectra. 
{\color{black}Some bins on the top has the value 1, this is because there are few DLAs with $\log{N_{\rm{HI}}}$ above 22 from Y1 mock. For example, there are just 2 DLAs in the bin with S/N 4 to 5 and $\log{N_{\rm{HI}}}$ above 22. The DLA finder detects them both and no FP samples, so the purity goes to 1. There are just 5 DLAs in the bin with S/N 3 to 4 and $\log{N_{\rm{HI}}}$ above 22. The DLA finder detects three of them and produce 2 FP samples, so the purity goes down to 0.6. That is why the purity for high column density DLAs changes a lot in different bins. } \label{fig:purity}}
\end{figure*}

\begin{figure*}
\plotone{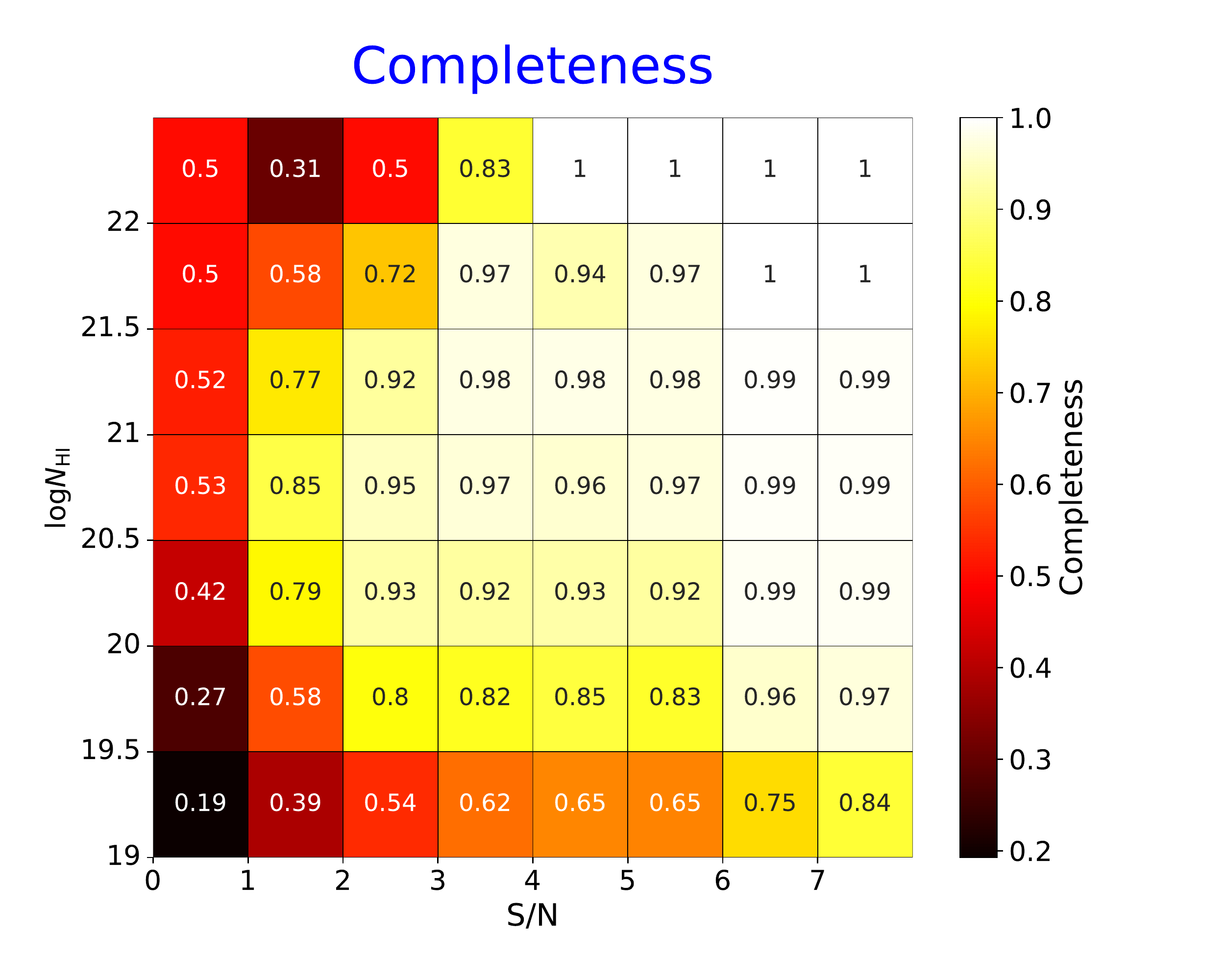}
\caption{Completeness results for different S/N levels and column densities using desi-Y1 mock spectra. This is the result choosing the critical value of $C_{\rm min}$ as 0.5. %The lower left bins are blank because we do not train our model to detect sub-DLAs in low S/N (S/N$<3$) spectra.
{\color{black}Some bins on the top has the value 1, this is because there are few DLAs with $\log{N_{\rm{HI}}}$ above 22 from Y1 mock. For example, there are just 2 DLAs in the bin with S/N 4 to 5 and $\log{N_{\rm{HI}}}$ above 22. The DLA finder detects them both, so the purity goes to 1. There are just 5 DLAs in the bin with S/N 3 to 4 and $\log{N_{\rm{HI}}}$ above 22. The DLA finder detects four of them and misses one DLA, so the purity goes down to 0.8. That is why the completeness for high column density DLAs changes a lot in different bins. } \label{fig:completeness}}
\end{figure*}

\begin{figure*}
\gridline{\fig{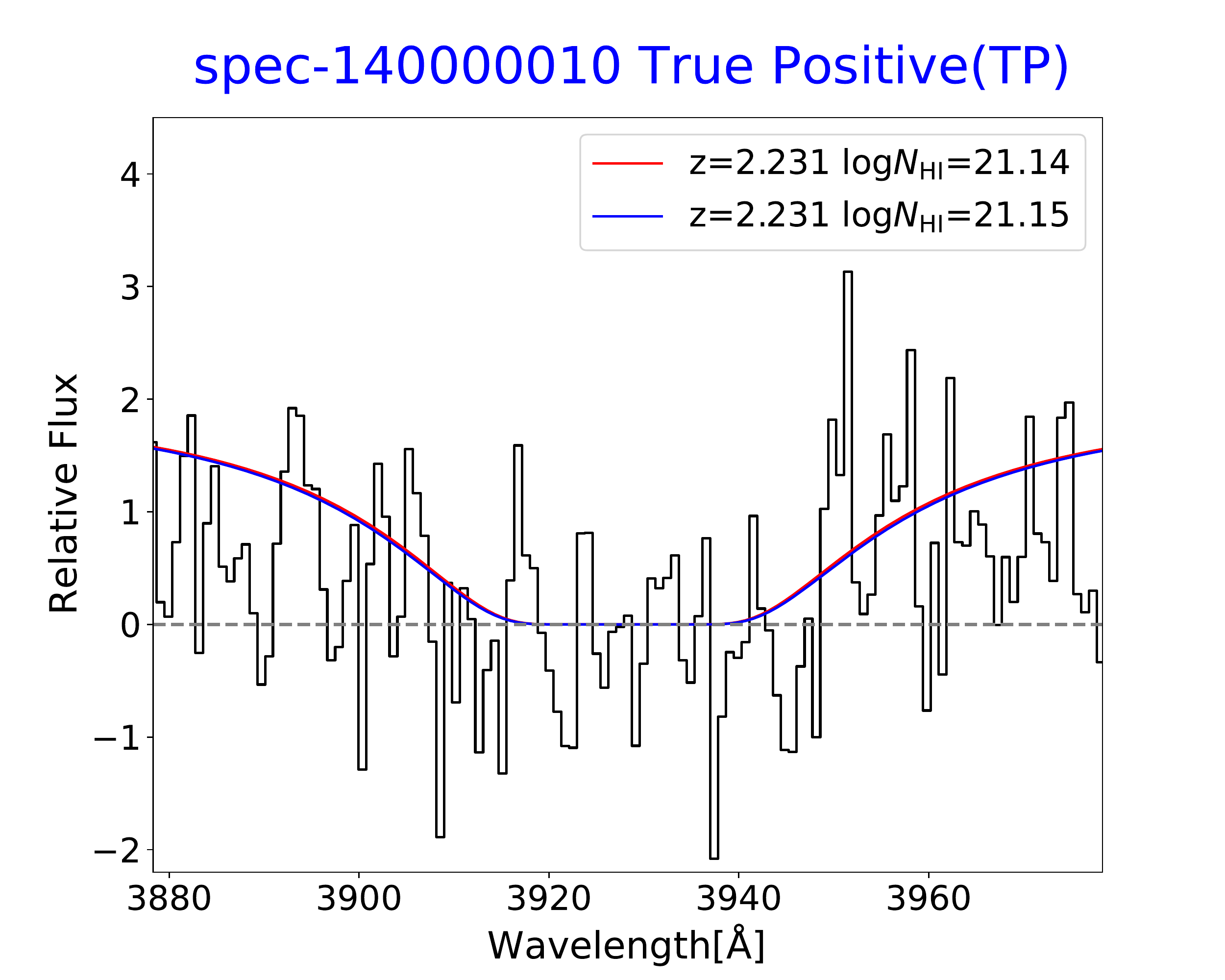}{0.5\textwidth}{(a)}
          \fig{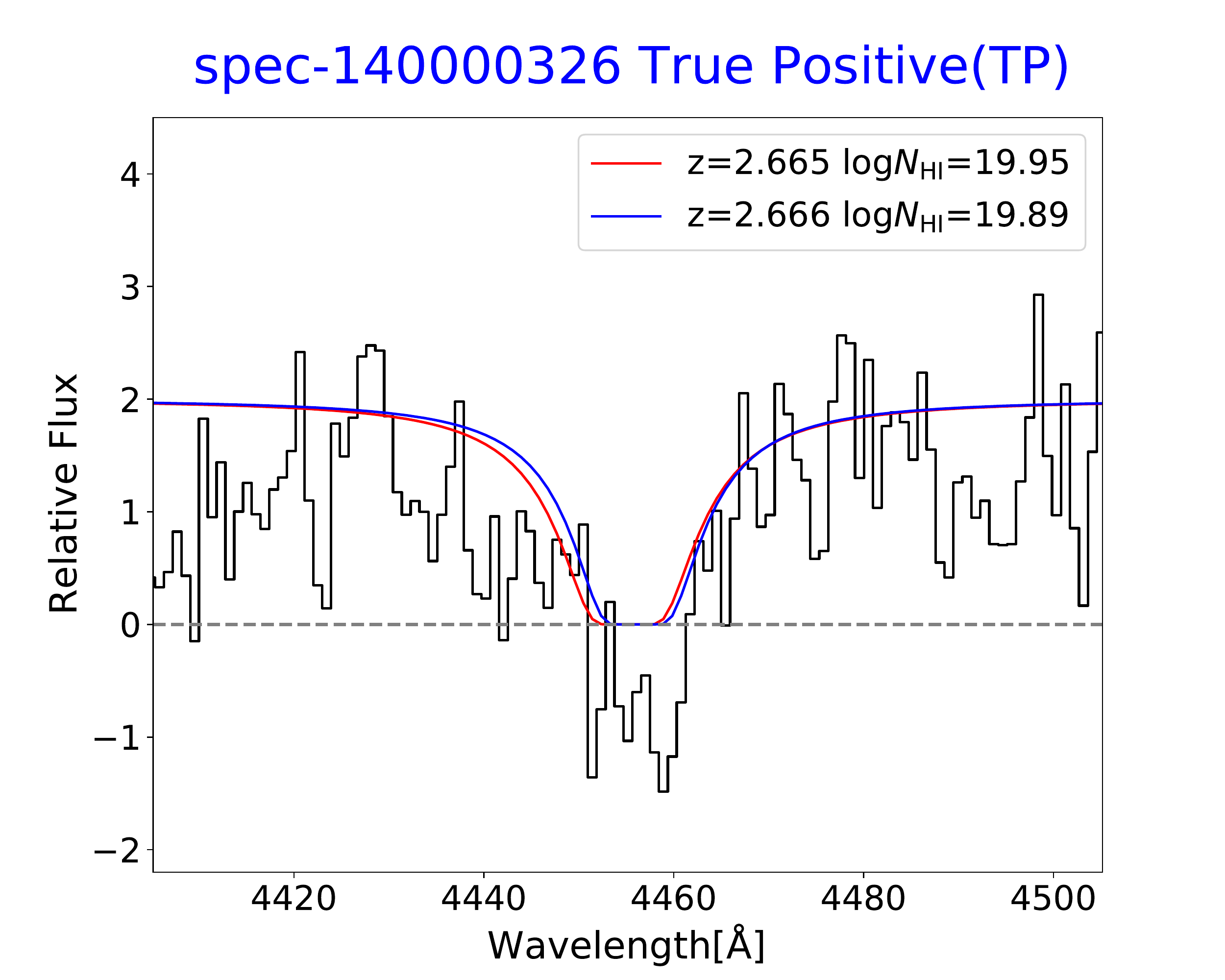}{0.5\textwidth}{(b)}
          }
\gridline{\fig{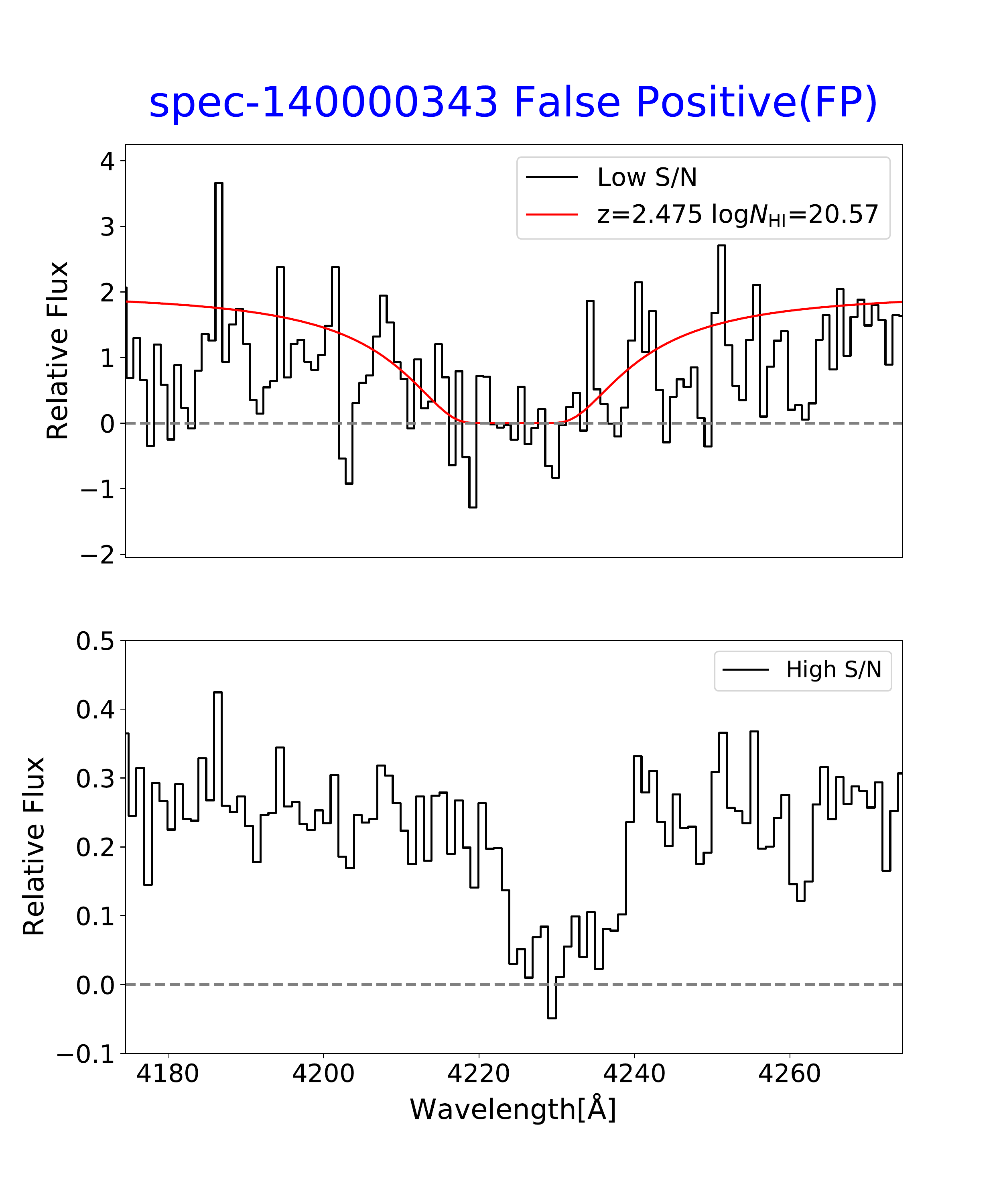}{0.5\textwidth}{(c)}
          \fig{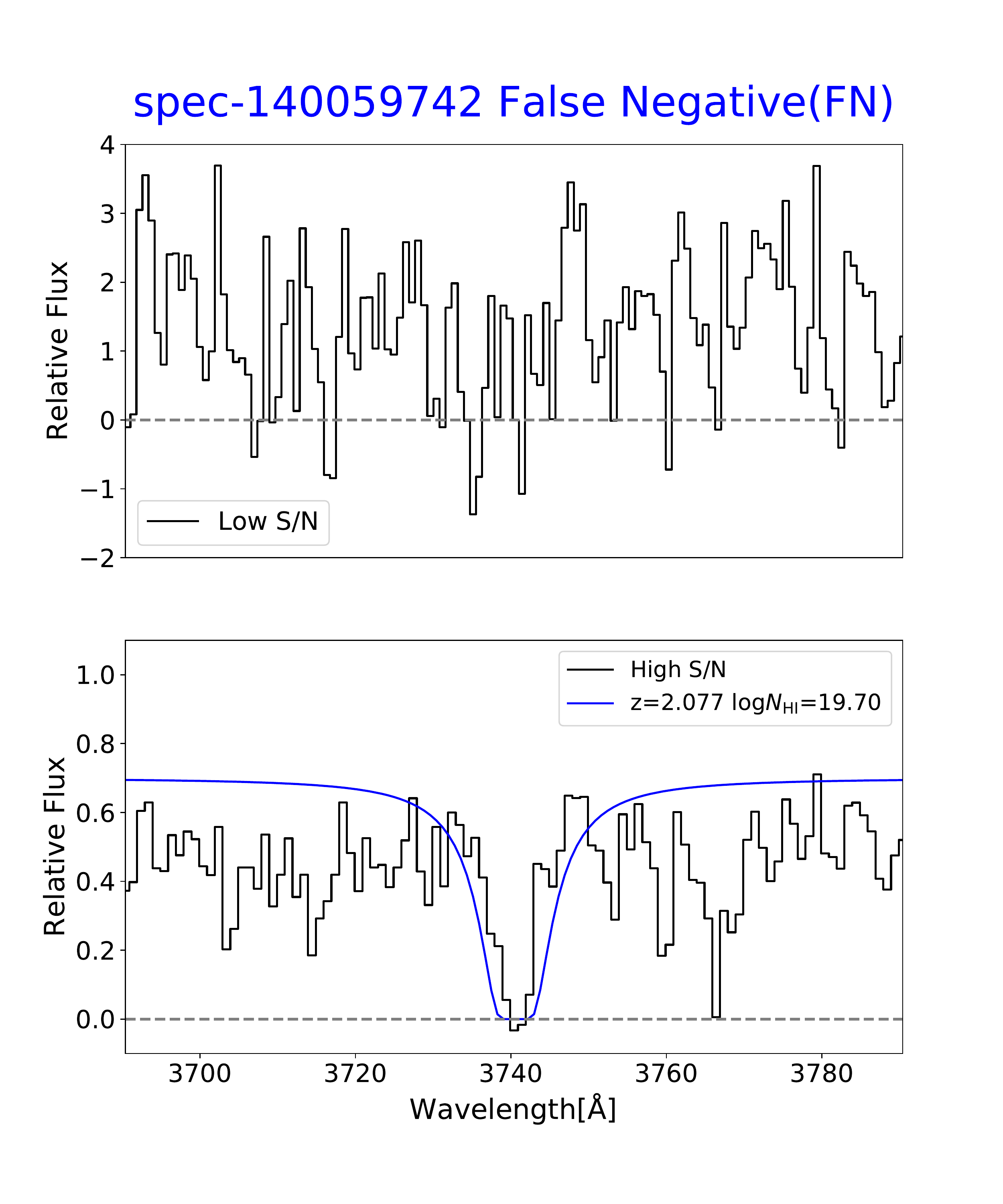}{0.5\textwidth}{(d)}
          }
\caption{validation samples: red lines are DLAs detected by our CNN model and blues lines are the DLAs in mock spectra. (a)(b) TP case, our model can detect DLAs with different column density levels. Even sub-DLAs can be characterized.  {\color{black}(c) FP case, the DLA finder identifies a DLA in this window but there is no such a DLA in the mock catalog. Lower panel of (c) is the same spectra as upper panel but with a higher S/N. It is clear there is no DLA if we check the high S/N spectra. But this is very difficult to identify even for the human being in upper panel of figure (c) because of the low S/N.  (d) FN sample, the DLA finder misses a sub-DLA in this wavelength range. Lower panel of (d) is the same spectra as upper panel but with higher S/N. This shows a missing sub-DLA with low S/N($<$2). The flux range for lower panel and upper panel of (c)(d) is quite different because a random array as the noise is inserted to get a lower S/N}}
%most of the missing DLAs are with low column density and low S/N($<$2)}. 
\label{fig:tpfpfn}
\end{figure*}

We have calculated the purity and completeness for different S/N and column densities. 
The results are shown in Figure \ref{fig:purity} and Figure \ref{fig:completeness}. 
For DLAs in spectra with S/N $>3$, our model can achieve both purity and completeness more than 90$\%$ for almost all column density levels. Although efforts have been made to optimize the DLA identification accuracy, 
FN and FP cases are inevitable. 
The majority of these occur in spectra
with low S/N ($<3$). 
As shown in Figure \ref{fig:tpfpfn}, these are
examples which are difficult to classify even for 
an expert. 

{\color{black}We also test our model on desiY1-0.14 mock spectra. This mock spectra contains both DLAs and BALs. We show the purity and completeness in Figure \ref{fig:bal-pc}. The completeness is still as good as shown in Figure \ref{fig:completeness}. The purity drops about 10$\%$ to 20$\%$ in different bins. This is because some BALs are identified as DLAs. Nevertheless, the DESI has a formal BAL catalog which will get rid of more than $98.6\%$ BALs \citet{2019ApJ...879...72G} from the catalog. Then, we can run DLA finder on the BAL-removed spectra. Therefore, we think that the purity result shown in Figure \ref{fig:purity} is still valid. }
%We do not show the purity and completeness for sub-DLAs at S/N $<3$ because the CNN model is not trained using these samples. 
%These sub-DLAs are very difficult to identify and they are not good training samples. 
%Thus, we do not use our CNN model to detect sub-DLAs using such low S/N spectra. 

%Although efforts have been made to optimize the DLA identification accuracy, 
%FN and FP cases are inevitable. 
%The majority of these occur in spectra
%with low S/N ($<3$). 
%As shown in Figure \ref{fig:tpfpfn}, these are
%examples which are difficult to classify even for 
%an expert. 

\subsection{Redshift Estimation} \label{subsec:Redshift Estimation}
According to the offset label predicted by CNN model,
we locate the central wavelength of DLAs.
This can be used to estimate the redshift of DLAs. 
The direct result for our model is the DLA location in every window.
We need to transfer this result to the location in the sightline. 
This procedure is similar in \citet{2018MNRAS.476.1151P}. 
We make the histogram of all the offset values, 
and a cluster of values at the center of a true DLA is expected. 
A confidence parameter for the whole spectra is further defined as the sum of the histogram over the nearest five pixels. 
After normalizing by the 9-pixel median filter, 
the maximum limit is set to one. 
Every detection with this confidence parameter above the critical value 
are considered as a DLA. 
Then we can calculate the redshift of DLAs according to the central wavelength. 

\begin{figure}
\plotone{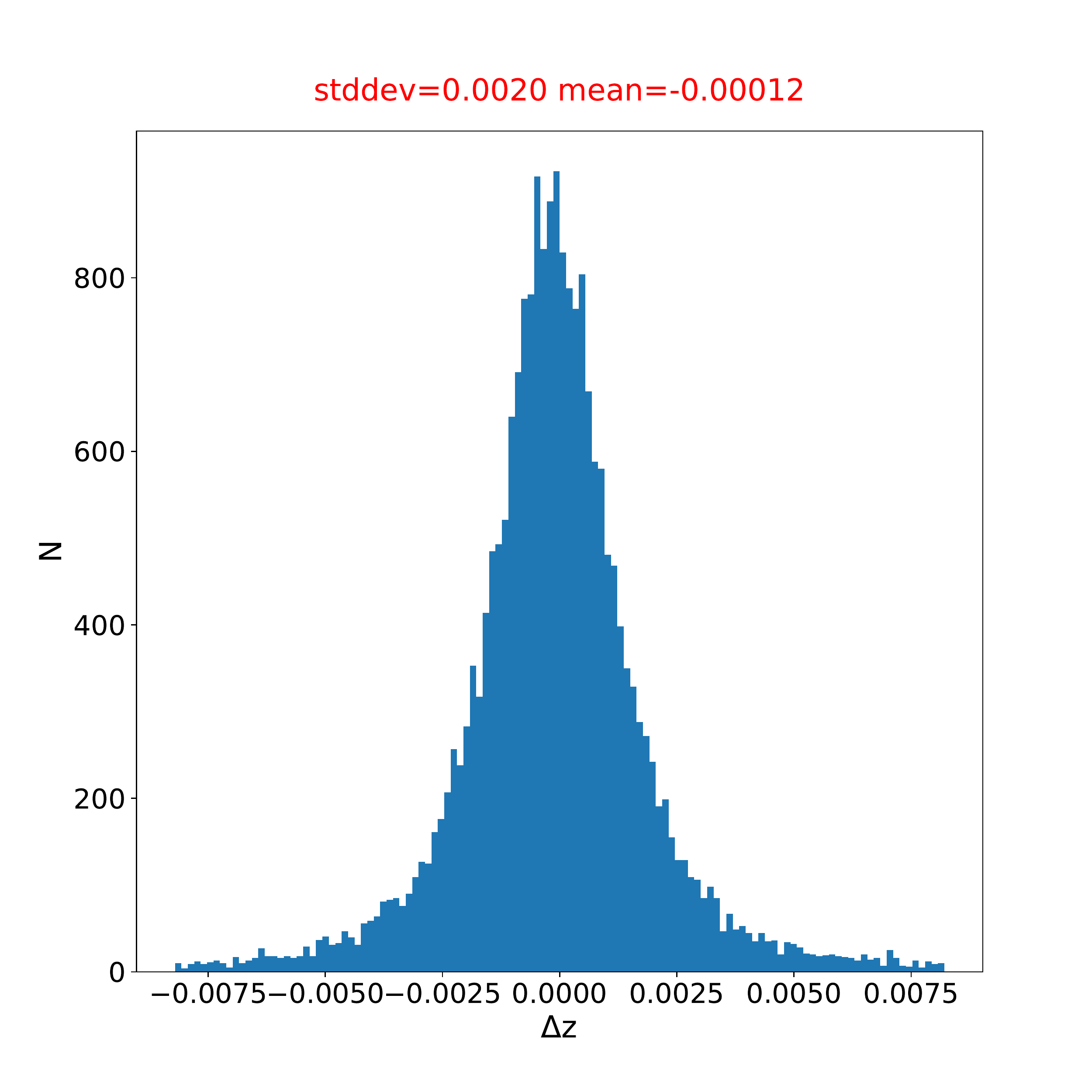}
\caption{Redshift estimation for the DLAs matched between the CNN model and the true value in mock spectra with $S/N > 3$ and $log(N_{\rm{HI}})> 20.0$. \label{fig:deltaz}}
\end{figure}

The difference of the redshift estimation compared to the true value (value in mock spectra) is shown in Figure \ref{fig:deltaz}. This result is for the spectra in 'desi-0.2-100' mock spectra. The mean value of the difference is -0.00012, and the standard deviation $\sigma(z)$ is 0.002. 
%This is showed in figure , 

\subsection{Column Density Estimation} \label{subsec:Column Density Estimation}
\begin{figure}
\plotone{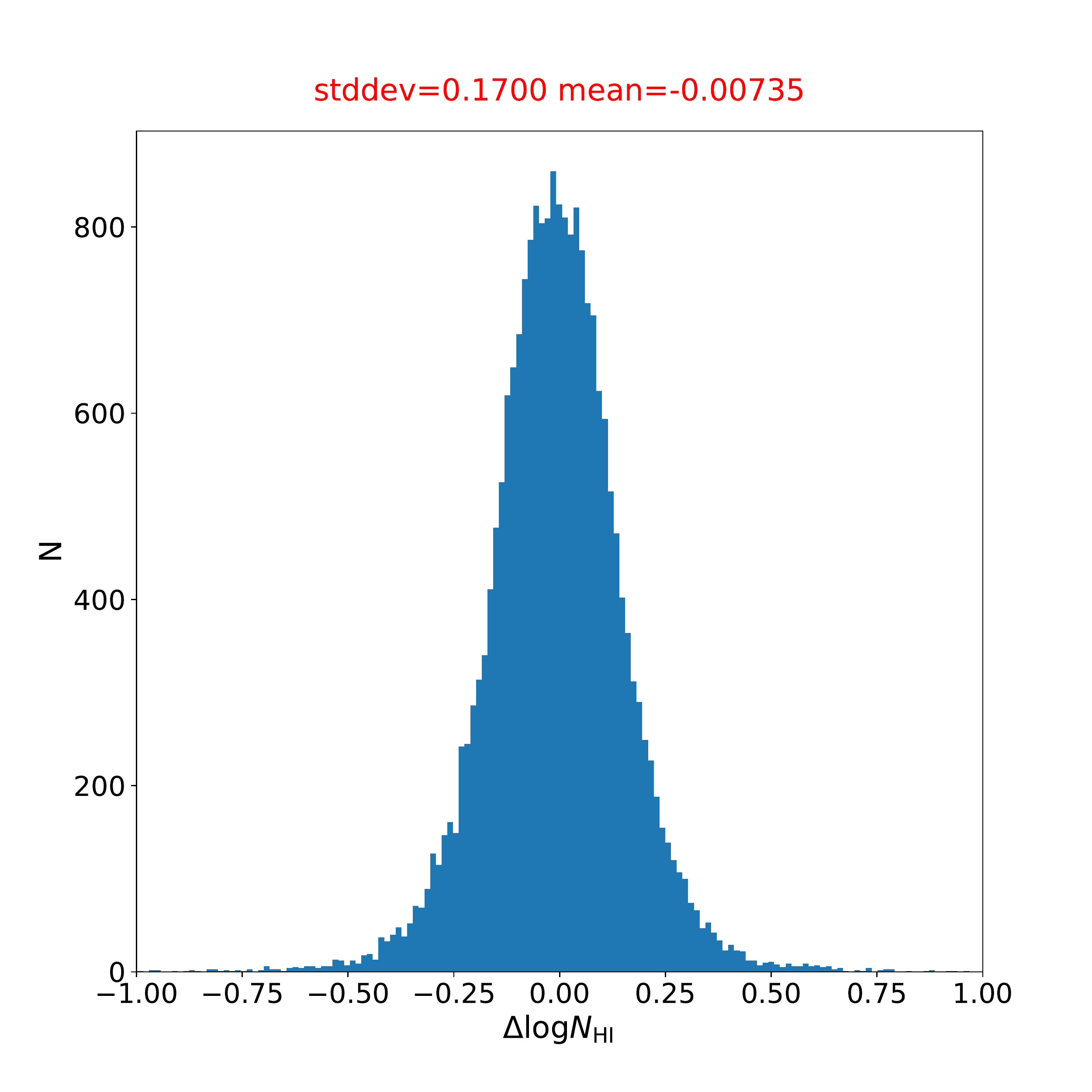}
\caption{The column density estimation for the DLAs matched between the CNN model and the true value in mock spectra with $S/N > 3$ and $log(N_{\rm{HI}})> 20.0$ . \label{fig:deltaNHI}}
\end{figure}

Our model can also give the estimation for the column density of DLAs. 
For every window of spectra,
we can get an estimated value of the column density. 
After locating the central wavelength of a DLA,
we can get the $N_{\rm{HI}}$ estimation results for the 40 pixels near the center,
and take the average value of these 40 $N_{\rm{HI}}$ estimate as the final estimate. 
The difference of the column density estimation compared to the true value is shown in Figure \ref{fig:deltaNHI}. This result is for the spectra in 'desi-0.2-100' mock spectra. 
The mean value of the difference is -0.007, and the standard deviation $\sigma(\log{N_{\rm{HI}}})$ is 0.17. 
%{\bf Also, we should summary the main results in Figure12 quantitatively here.}

%\section{Discussion} \label{discussion}
\section{Comparison with Gaussian Process Model} \label{sec:Gaussian process}

Recent advances in Gaussian Process (GP) model facilitated the investigation of detecting DLAs from SDSS\citep{2017MNRAS.472.1850G,2020MNRAS.496.5436H,2021arXiv210310964H}.   \citet{2021arXiv210310964H} presented a DLA catalog from SDSS DR16Q, with an improved GP model. Currently, it might be the most solid approach to compare the performance between CNN model and GP model using DESI mock spectra with a given DLA catalog.

{\color{black}Here we briefly introduce their GP model based on the Bayesian model selection in \citet{2020MNRAS.496.5436H}.
A set of models ${\mathcal{M}_{i}}$ are developed, including the model without DLAs ($\mathcal{M}_{\rm{\neg DLA}}$), the models with 4 DLAs (${\mathcal{M}_{\rm{DLA(i)}}}_{i=1}^{4}$), and the model with sub-DLAs ($\mathcal{M}_{\rm{sub}}$).
These models use Gaussian Processes to describe the QSO emission function, a qso’s true emission spectrum $f(\lambda)$. Then they add the instrumental noise and absorption due to the intervening intergalactic medium (IGM) to obtain the observed flux as a function $y(\lambda)$.
With a given spectroscopic sightline $\mathcal{D}$, they can evaluate the posterior probability of these model based on Bayes’s rule:
\begin{equation}
    Pr(\mathcal{M}\mid \mathcal{D})=\frac{p(\mathcal{D}\mid \mathcal{M})Pr(\mathcal{M})}{\sum_i p(\mathcal{D}\mid \mathcal{M}_{i})Pr(\mathcal{M}_{i})}
\end{equation}
where $p(\mathcal{D}\mid \mathcal{M})$ is the model evidence of the QSO spectrum $\mathcal{D}$ given model $\mathcal{M}$, $Pr(\mathcal{M})$ is the prior probability of
model $\mathcal{M}$, and the denominator on the right-hand-side is the
sum of posterior probabilities of all models in consideration.}

Following the pipeline described in \citet{2020MNRAS.496.5436H}, we firstly retrained the null model $M_{\rm{\neg DLA}}$ using 70255 spectra without DLAs in the 'desiY1-0.2-DLA' mock. 
{\color{black}Then we extend the null model $M_{\rm{\neg DLA}}$ to a model with k intervening DLAs, $\mathcal{M}_{\rm{DLA(k)}}$(k up to 4).}
The model prior and model evidence for {\color{black}these models} are approximated by using the 'desiY1-0.2-DLA' mock DLA catalog.  
Applying the new GP model, we obtain a DLA catalog of 248512 sightlines in the 'desiY1-0.2-DLA' mock. 
{\color{black}Note that the default Voigt profile they use in \citet{2020MNRAS.496.5436H} includes Ly$\alpha$,
Ly$\beta$, and Ly$\gamma$ absorption, \textcolor{black}{but we set the number of absorption lines $num\_lines$ to one since this mock only contains Ly$\alpha$ absorption lines.\footnote{ Note that the modification of this parameter may limit the performance of the GP model, which \citet{2021arXiv210310964H} claimed that GP model performs better in the $Ly\beta$ forest than the CNN model. But this is not discussed in this manuscript.}} Also we modify the parameters about the mininum distance between DLAs $min\_z\_separation$ to zero since we want to identify very close overlapping DLAs.} All codes related to GP are available at \url{https://github.com/zoujiaqi99/GP_DLA_DESI}.

\begin{figure}
\plotone{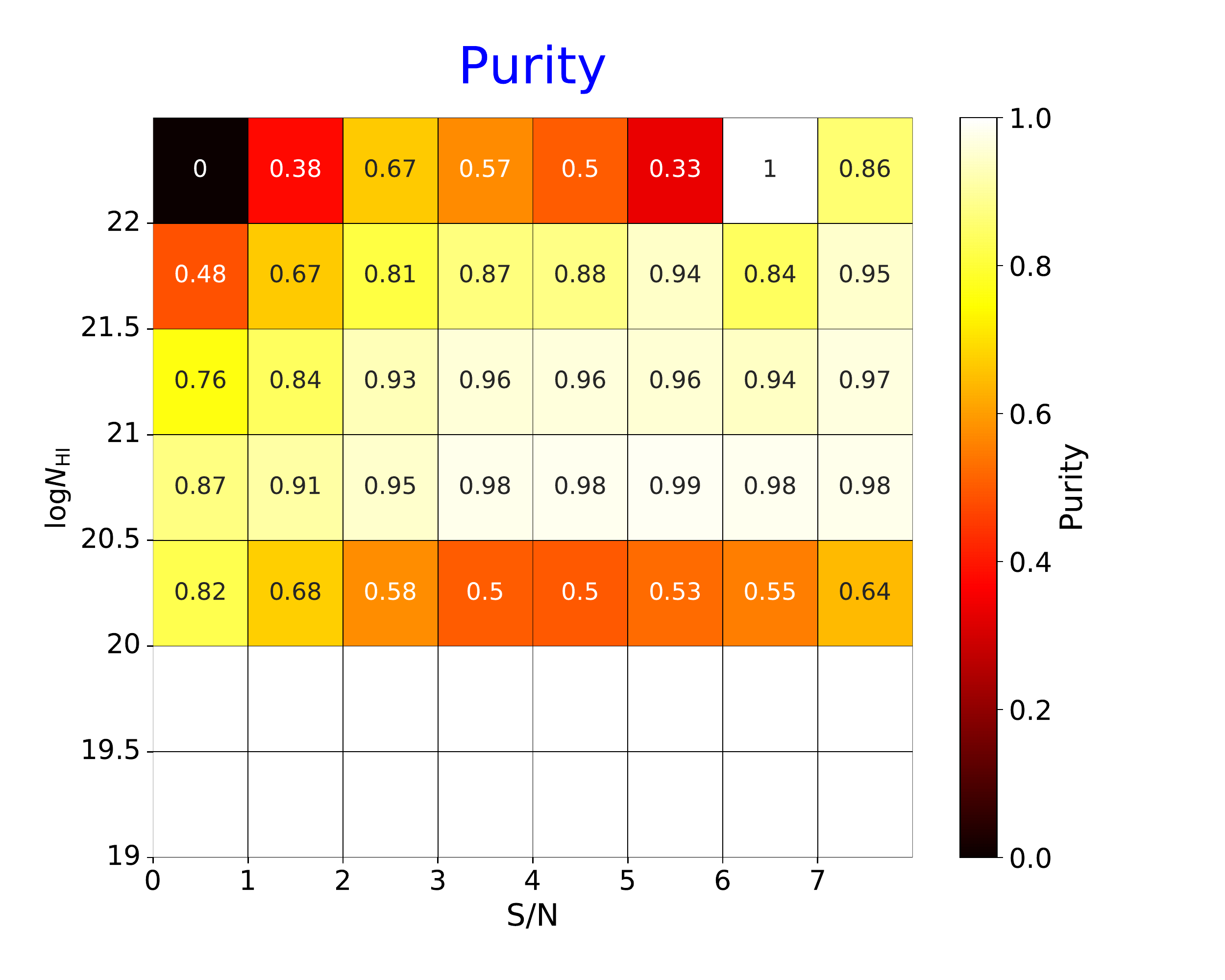}
\caption{GP purity results for different S/N levels and column densities using desi-Y1 mock spectra. Here we only select absorbers with  $N_{\rm{HI}}> 20.0$ to provide a simple comparison. Because the current GP model is well developed only using $N_{\rm{HI}}> 20.0$ absorbers.  \label{fig:gp purity}}
\end{figure}

\begin{figure}
\plotone{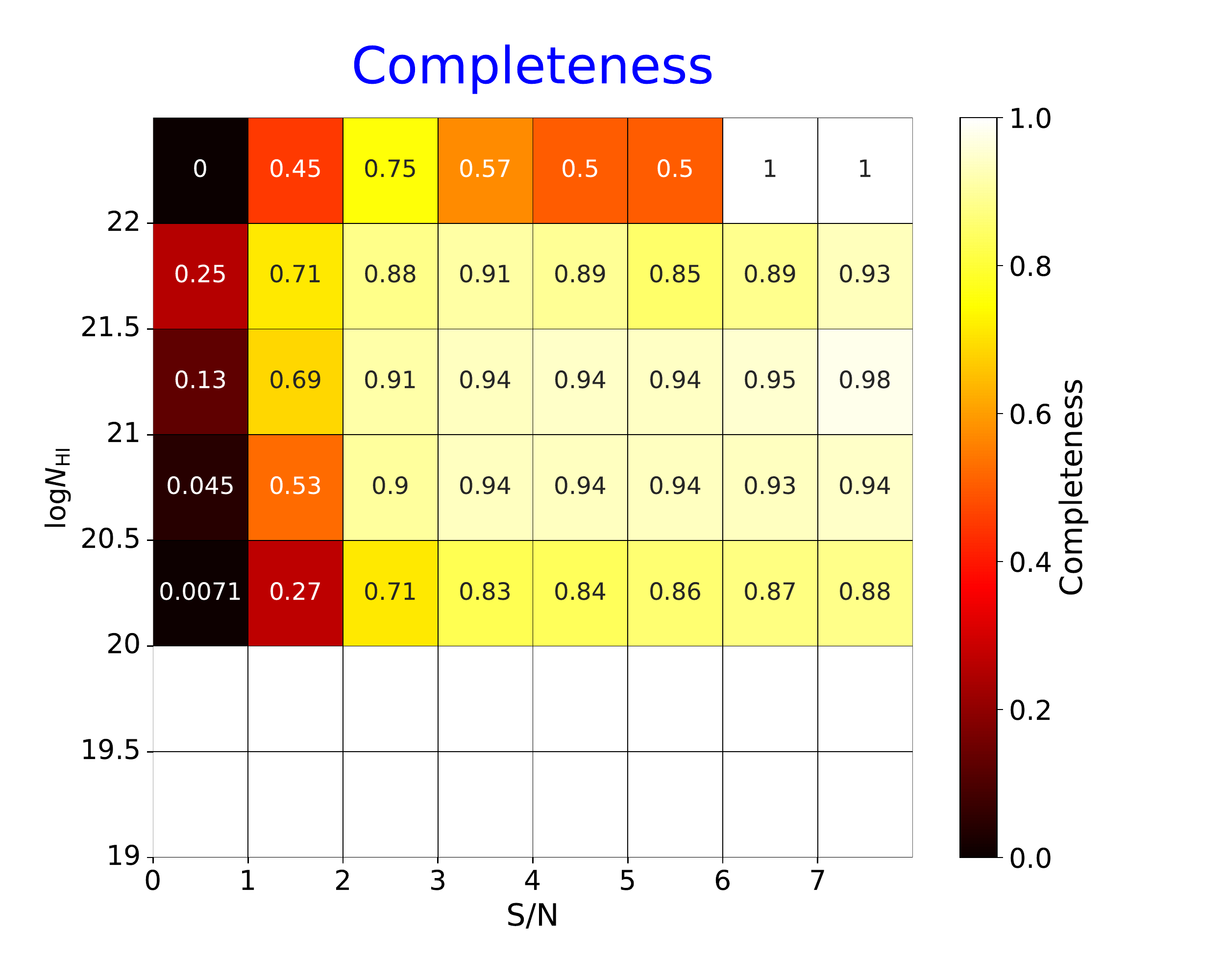}
\caption{GP completeness results for different S/N levels and column densities using desi-Y1 mock spectra. Here we only select absorbers with  $N_{\rm{HI}}> 20.0$ to provide a simple comparison. Because the current GP model is well developed only using $N_{\rm{HI}}> 20.0$ absorbers.  \label{fig:gp completeness}}
\end{figure}

With $S/N > 3$ and $log(N_{\rm{HI}})> 20.0$, there are 18613 real DLAs in the mock 
DLA catalog. 17571 DLAs are predicted by the CNN model while 23212 DLAs are predicted by the GP model.
As discussed in Section \ref{sec:validation}, we also present the purity and completeness for S/N and column densities level as shown in Figure \ref{fig:gp purity} and Figure \ref{fig:gp completeness}. For GP model, completeness and purity are both greater than 88$\%$ for $S/N > 3$. \textcolor{black}{Note that our CNN model can measure the redshift and column density with $log(N_{\rm{HI}})> 19.3$. 
The GP model we used provides model posterior probability of whether the sightline containing absorbers with log(NHI) $<$ 20.0, but does not save the exact redshifts or column densities.  
%The current GP model can only give the model posterior probability of whether the sightline containing absorbers with $log(N_{\rm{HI}})< 20.0$, but cannot provide the exact redshifts or column densities. 
This is due to the difference in training sets between the CNN and GP methods. We fairly compare the two models under the same conditions in this manuscript. However, the DESI mock spectra allow us to build datasets with low $N_{\rm{HI}}$ absorbers to re-train the GP model in order to decrease its minimum to $log(N_{rm{HI}})=19.3$. }

\begin{figure}
\plotone{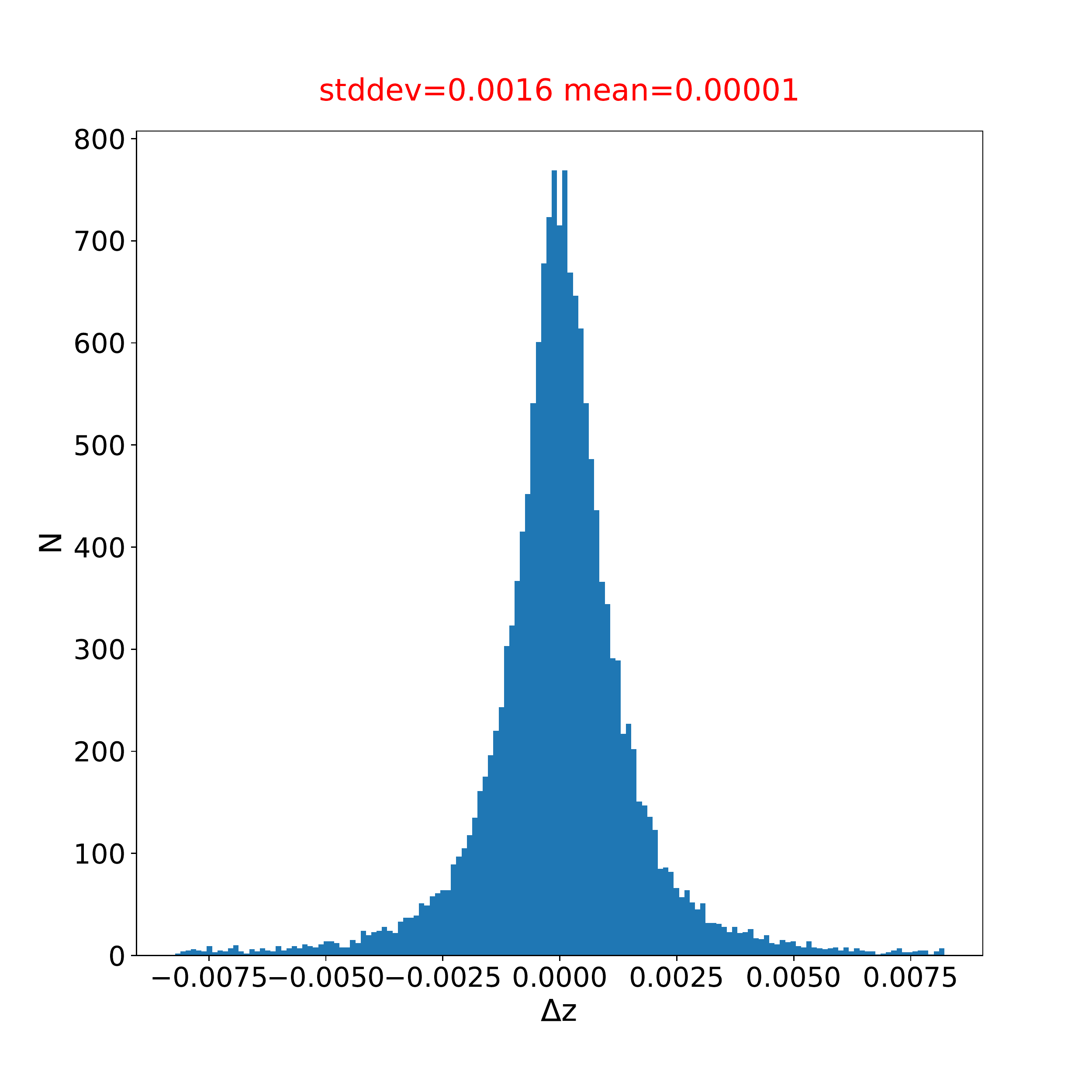}
\caption{The redshift estimation for the DLAs matched between the Gaussian Process model and the true value in mock spectra with $S/N>3$ and $\log{N_{\rm{HI}}}>20.0$ \label{fig:deltaz_GP}}
\end{figure}

\begin{figure}
\plotone{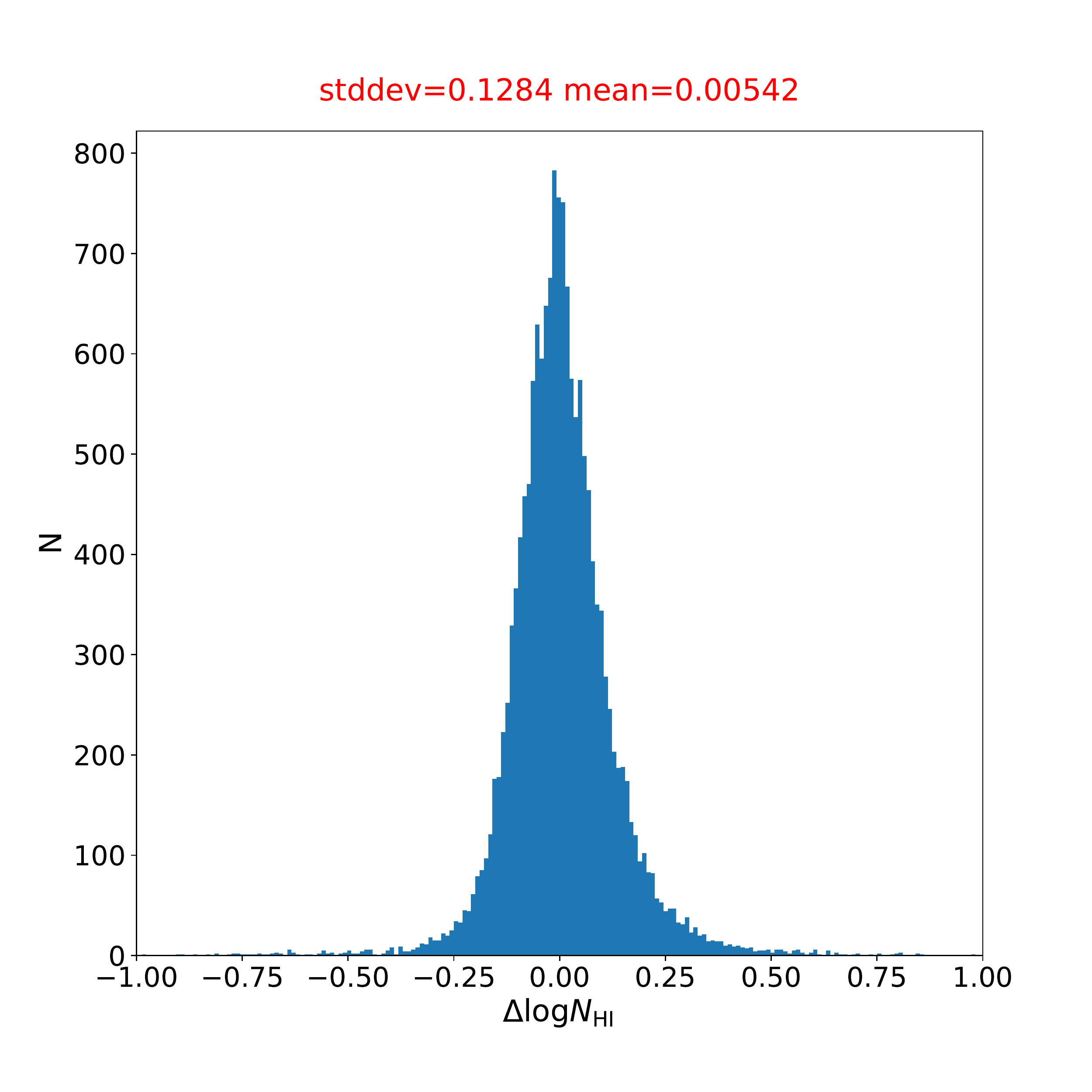}
\caption{The column density estimation for the DLAs matched between the Gaussian Process model and the true value in mock spectra with $S/N>3$ and $\log{N_{\rm{HI}}}>20.0$ \label{fig:deltaNHI_GP}}
\end{figure}

{\color{black}Figure \ref{fig:deltaz_GP} and \ref{fig:deltaNHI_GP} show histograms of the offsets in
redshift and $N_{\rm{HI}}$ between the GP model’s predictions and
real values in the mock catalog. The mean redshift
offset is 0.00001 with a standard deviation of 0.0016. The
mean log column density offset is $\Delta log(N_{\rm{HI}})= 0.005$ with
standard deviation of 0.13 dex.}

\begin{figure}
\plotone{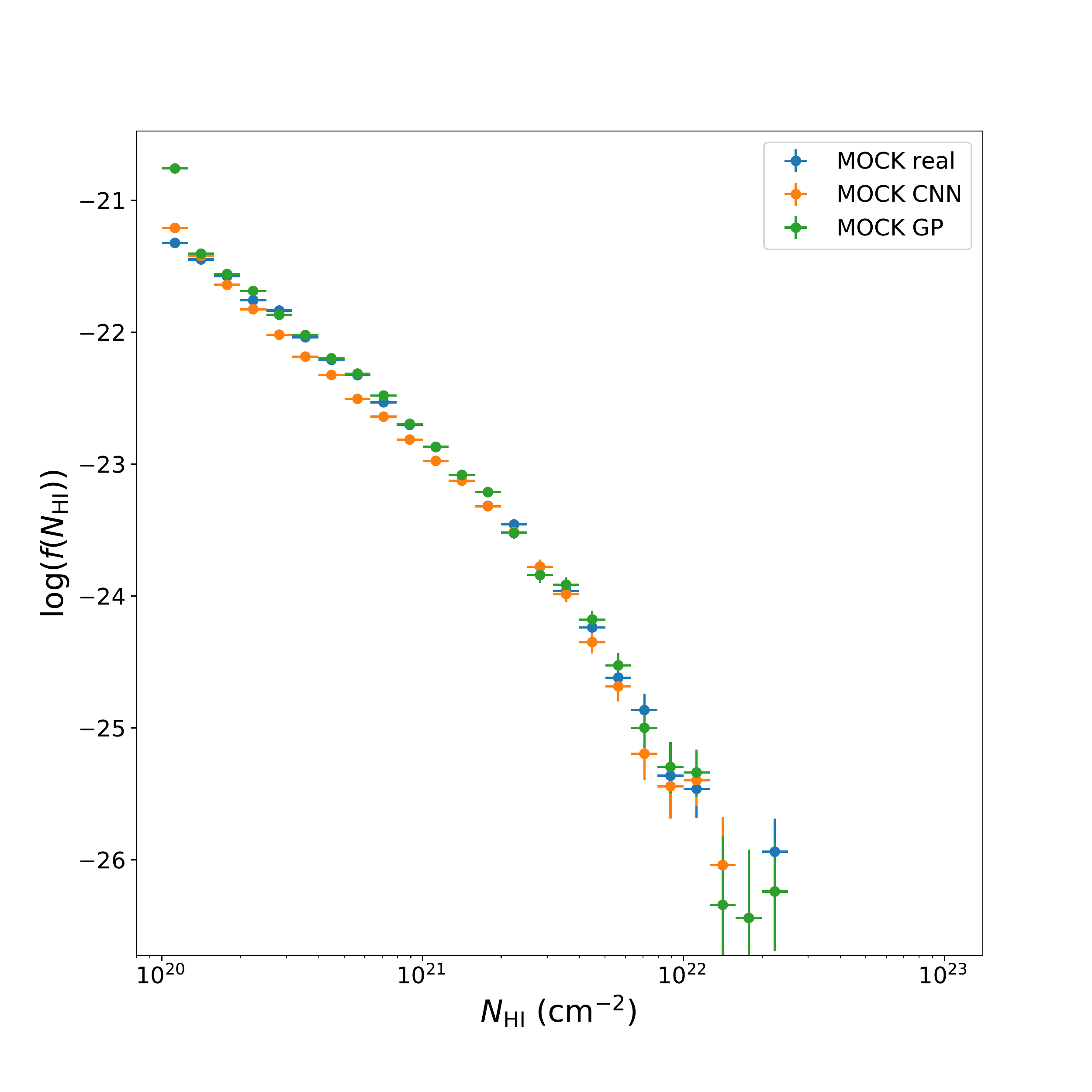}
\caption{ The CDDF from both CNN DLA catalog and GP DLA catalog in comparison to the real mock DLA catalog with $z<3.8$. The error bars in y-axis represent the 68$\%$ confidence limits.\label{fig:CDDF}}
\end{figure}

 We also present the column density distribution function (CDDF) in Figure \ref{fig:CDDF}. This figure contains both CNN DLA catalog and GP DLA catalog in comparison to the real mock DLA catalog with $z<3.8$. The distribution of both models does not significantly differ from that of the real catalog. Due to the more accurate $N_{\rm{HI}}$ estimation, the GP model is in a better agreement with the real catalogue except for the Monte Carlo sampling boundary($log(N_{\rm{HI}}) =20.0$), where induces the over-detection of absorbers with $log(N_{\rm{HI}}) <20.0$. \citet{2020MNRAS.496.5436H} showed the CDDF and concluded that the previous CNN model \citep{2018MNRAS.476.1151P} is failing to detect $> 60\%$ of DLAs with $log(N_{\rm{HI}}) > 21$. This problem does not exist in our results. 
\textcolor{black}{The lack of high column density absorbers 
in the Parks catalogue may be due to a lack of high column density systems in their training set. } %Their conclusion may be due to the fact that only a small DLA sample with $p_{DLA}>0.9$ were used in their analysis. }

%\begin{deluxetable}{ccccccc}
%\tablecaption{Best fitting parameters for GP\label{tab:gp bao}}
%\tablehead{
%\colhead{parameter} & \colhead{DLA GP} &\colhead{DLA mock}   &\colhead{difference}
%}
%\startdata
%ap  &  0.994  & 0.983 & 1.11$\%$\\
%\sigma & 0.021 & 0.021 & \\
%at &  1.023 & 1.015 & 0.79$\%$\\
%\sigma & 0.021 & 0.021 & \\
%$\chi^{2}/DOF$ & 1.168 & 1.096 & \\
%\enddata
%
%\end{deluxetable}

Compare the performance of the same mock, we find that the CNN model performs better in purity and completeness while the GP model has a more accurate estimation of column density. In terms of BAO measurement {\color{black}using DESI Y1 mock}, the CNN model is more effective as described in Section \ref{sec:BAO}.
Besides, GP model takes 8 to 10 times longer than CNN model to predict the same dataset. 
{\color{black}Consequently, a combined DLA catalog that takes the best of both models might be a better choice for DESI real data. CNN can be mainly used to detect DLAs and estimate redshift because of the higher completeness and purity. GP can be applied to further improve the column density estimate. The DLAs detected by GP are an important part to complete the DLA catalog. Besides, we can also provide a DLA catalog which only contains DLAs detected by both CNN and GP. The DLAs detected by both algorithms  maybe a smaller sample but with high confidence.  }

\section{BAO fitting analysis} \label{sec:BAO}

\subsection{BAO fitting procedure} \label{subsec:BAO1}

After the DLA catalogs are generated,
we tested the influence of different DLA catalogs on the measurement of BAO fitting. 
{\color{black}Two different DLAs catalogs are generated. 
One catalog contains all the DLAs in the mock spectra, 
we will take this catalog as the real DLA catalog. }
Another catalog only contains the DLAs detected by our CNN model. 
The BAO fitting is conducted using the Ly$\alpha$-QSO cross correlation and Ly$\alpha$-Ly$\alpha$ auto correlation. 
According to the definition of \citet{2020ApJ...901..153D}, the flux-transmission field is 

\begin{equation}
\delta_{q}(\lambda)=\frac{f_{q}(\lambda)}{F(\lambda)C_{q}(\lambda)}-1
%\delta_q(\lambda)=\frac{f_{q}(\lamda)}{F(\lamda)C_{q}(\lamda)}-1
\label{equ:fluxfield}
\end{equation}

Where $f_{q}(\lambda)$ is the observed flux, $F(\lambda)C_{q}(\lambda)$ is the mean expected flux. 
The BAO fitting procedure is followed by the pipeline in \citet{2020ApJ...901..153D}. 
{\color{black}For the Ly$\alpha$-QSO cross correlation and Ly$\alpha$-Ly$\alpha$ auto correlation, the first step is to do the continuum fitting. In this step, we obtain the flux-transmission field  from the observed flux and the mean expected flux. DLAs should be masked in this step. }

%, the procedure is
%\begin{enumerate}
%\item Continuum fitting and delta field: obtain the flux-transmission field  from the observed flux and the mean expected flux. {\color{black}DLAs should be masked in this step. }
%\item Measuring correlations: It is defined to be the weighted mean of the {\color{black}fluctuation of the transmission} at a given separation from a QSO.
%\item Measuring the distortion matrix: measure the distortion matrix, it is used to model the effect of continuum fitting in the correlation function.
%\item Measuring the metal matrix (optional): Metal lines can contaminate the Ly$\alpha$ forest and thus this should be considered. BAO fitter has a parameter of $picca\_metal\_xdmat$ which taking into account. 
%\item Fitting BAO: Use the results above to get the BAO signal. 
%\end{enumerate}
%The procedure for getting the Ly$\alpha$-Ly$\alpha$ auto correlation is similar to these steps. 

The next steps and the meaning of parameters are described in details in \citet{2020ApJ...901..153D}. 
%{\color{black}DLAs should be masked for the continuum fitting step\citep{2020ApJ...901..153D}. 
{\color{black}The pipeline we use for these steps is called 'PICCA' (\url{https://github.com/igmhub/picca}), which was developed by the eBOSS Ly$\alpha$ working group \citep{2020ApJ...901..153D}. This pipeline can mask DLAs for the BAO fitting if we provide a DLA catalog as an input. }

{\color{black}In this fitting, we have set the following parameters in PICCA as free parameter to fit. The high column density systems(HCD) are also considered in the fitting:
\begin{enumerate}
\item $\alpha_\parallel$, $\alpha_\perp$: BAO-peak position parameters. 
\item $b_{Ly\alpha}$, $\beta_{Ly\alpha}$: bias parameters for $Ly\alpha$ absorption.   
\item $b_{HCD}$, $\beta_{HCD}$: bias parameters for HCD systems.  
\end{enumerate}
Other parameters in PICCA has been set as fixed parameter. } 
%The complete parameter setting list can be found in the appendix. 

\subsection{BAO fitting results} \label{subsec:BAO2}

We have used the 'desiY1-0.2-DLA' mock spectra to do the BAO fitting analysis. 
{\color{black} This mock spectra contains DLAs and HCDs but do not have metal components. }

\begin{figure*}
\gridline{\fig{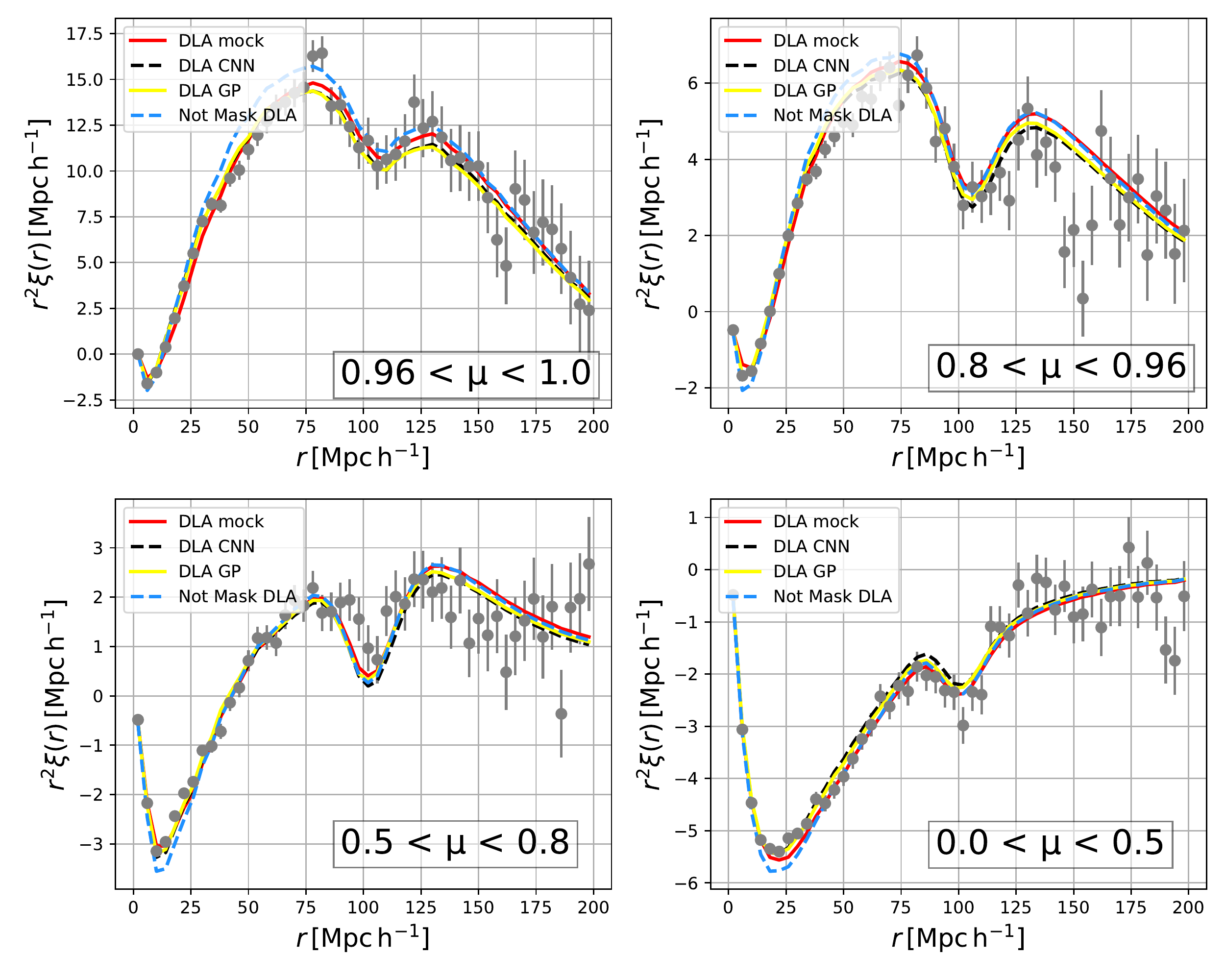}{0.5\textwidth}{(a)}
          \fig{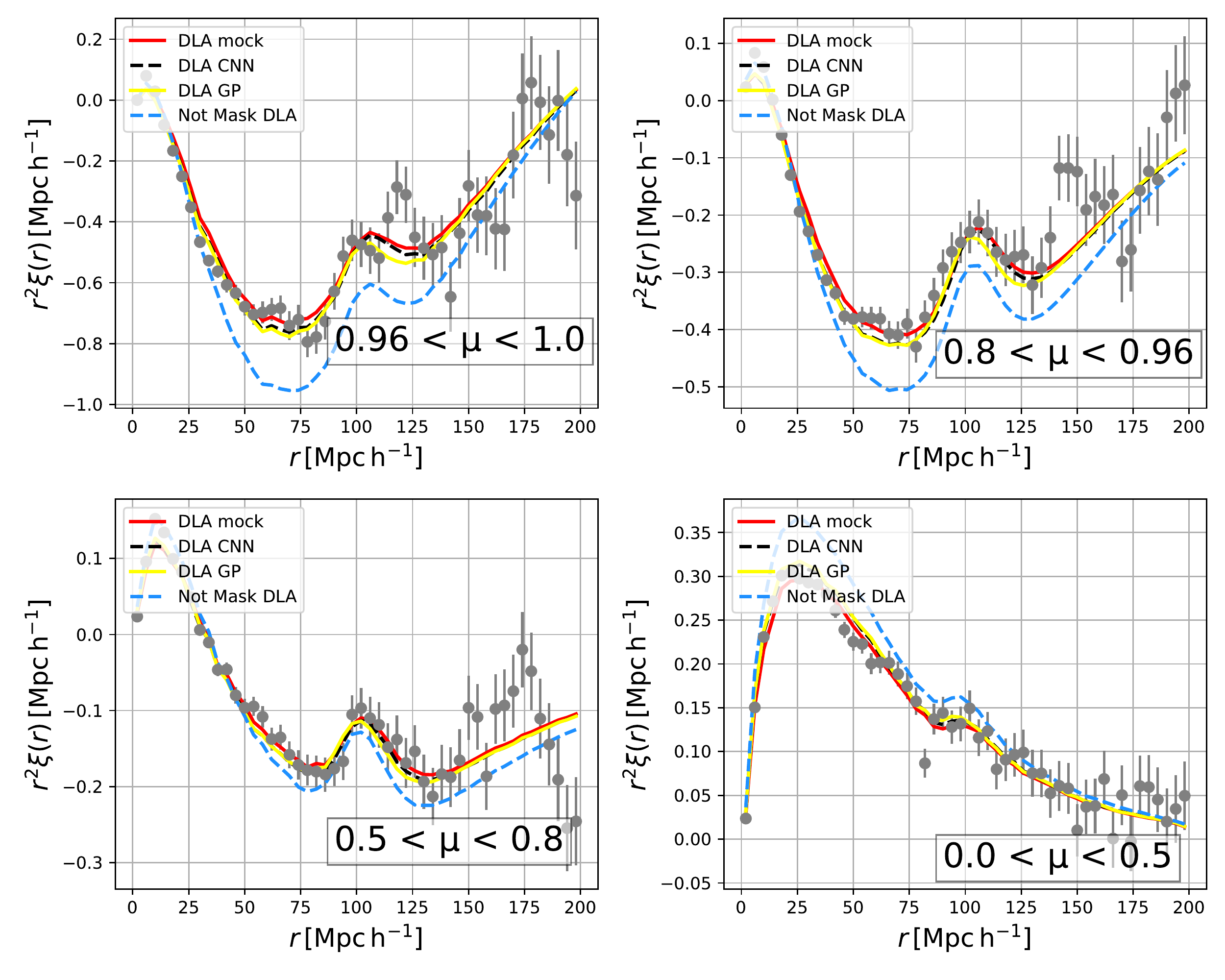}{0.5\textwidth}{(b)}
          }
%\gridline{\fig{figures/dla_auto_hcd.pdf}{0.8\textwidth}{(b)}
%          }
\caption{{\color{black}BAO fitting result:(a)the QSO-Lya cross-correlation (b) Lya-Lya auto-correlation. The red line is the best-fit model by masking the DLAs in the mock catalog. The black dashed line is the best-fit model by masking DLAs detected by our CNN model. The blue dashed line is the best-fit model if we do not mask any DLA. The yellow dashed line is the best-fit model by masking DLAs detected by Gaussian Process. The grey points are the data points from the analyse by masking the DLAs in mock spectra. }}
%The data points from two different fitting is hard to distinguish so we just plot the data point from one fitting. }
\label{fig:bao}
\end{figure*}

This mock spectra contains 212238
sightlines with 36212 DLAs and 43909 sub-DLAs. 
% 66058 sightlines with DLA
Our CNN model has detected 43530 DLA candidates and 19687 sub-DLAs. 
%Our CNN model has detected 38356 DLA candidates and 18903 sub-DLAs.
{\color{black} 38410 DLA candidates are detected by GP. The sub-DLAs in mock catalog are added to GP catalog when we do the BAO fitting. }
FP and FN samples are inevitable from the lower 
S/N spectra ($<3$). 
%Two different DLAs catalogs are generated. 
%One catalog contains all the DLAs in the mock spectra, 
%we will take this catalog as the real DLA catalog. 
%Another catalog only contains the DLAs detected by our CNN model. 
We have conducted the Ly$\alpha$-QSO cross correlation and Ly$\alpha$-Ly$\alpha$ auto correlation by masking these two different DLA catalogs. We also plot the fitting result if we do not mask any DLAs and mask the DLA catalog generated by Gaussian Process (GP) which were {\color{black}described} in Section \ref{sec:Gaussian process}. 
The results are shown in Figure \ref{fig:bao}. 

After masking DLAs, the reduced $\chi^{2}$ of the fitting decreases from 1.136 to 1.081. 
Impact of DLAs on BAO fitting is obvious in the auto-correlation fitting. 
The BAO fitting result from the mock DLAs catalog was set as
the ground-truth. 
{\color{black}
Two parameters describing the position of BAO peak \citet{2013A&A...552A..96B}, $\alpha_\parallel$ and $\alpha_\perp$ are estimated to quantify the fitting results. }
The best fitting parameters are shown in Table \ref{tab:bao parameter}. {\color{black}This table also contains the standard deviation for $\alpha_\parallel$, $\alpha_\perp$ and the reduced $\chi^{2}$ in two different fittings. }
%\ref{tab:cross parameter} and \ref{tab:auto parameter}. 
The difference of the results can be quantified by the following equations:

\begin{deluxetable}{ccccccc}
\tablecaption{\color{black}{Best fitting parameters \label{tab:bao parameter}}}
\tablehead{
\colhead{parameters} & \colhead{DLA mock} &\colhead{DLA CNN} &\colhead{DLA GP} &\colhead{No Mask}
}
\startdata
$\alpha_\parallel$  &  0.981  & 0.983 & 0.977 & 0.992\\
$\sigma$ & 0.0173 & 0.0174 & 0.0195 & 0.0212\\
$d_{\alpha_\parallel}$ &  & 0.16$\%$ & 0.41$\%$ & 1.12$\%$\\
ratio & & 0.116 & 0.209 & 0.636\\
$\alpha_\perp$ &  1.019 & 1.025 & 1.025 & 1.028\\
$\sigma$ & 0.0172 & 0.0177 & 0.0194 & 0.0227\\
difference &  & 0.61$\%$ & 0.61$\%$ & 0.94$\%$\\
ratio & & 0.349 & 0.304 & 0.523\\
$\chi^{2}/DOF$ & 1.081 & 1.091 & 1.098 & 1.136\\
%\caption{Shown are the values and standard deviation for ap, at and the reduced \chi^{2} in two different fittings}
\enddata

\end{deluxetable}

%\begin{deluxetable}{ccccccc}
%\tablecaption{Best fitting parameters \label{tab:bao parameter}}
%\tablehead{
%\colhead{parameters} & \colhead{DLA mock} &\colhead{DLA CNN} &\colhead{DLA GP} &\colhead{No Mask}
%}
%\startdata
%$\alpha_\parallel$  &  0.983  & 0.992 & 0.994 & 0.997\\
%\sigma & 0.0213 & 0.0217 & 0.0207 & 0.0208\\
%difference &  & 0.91$\%$ & 1.11$\%$ & 1.42$\%$\\
%$\alpha_\perp$ &  1.015 & 1.020 & 1.023 & 1.025\\
%\sigma & 0.0205 & 0.0212 & 0.0211 & 0.0230\\
%difference &  & 0.49$\%$ & 0.79$\%$ & 0.98$\%$\\
%$\chi^{2}/DOF$ & 1.096 & 1.071 & 1.168 & 1.164\\
%\caption{Shown are the values and standard deviation for ap, at and the reduced \chi^{2} in two different fittings}
%\enddata

%\end{deluxetable}

%The difference of the results can be quantified by the following equations:
\begin{equation}
d_{\alpha_\parallel}=\frac{|\alpha_{\parallel}(mock)-\alpha_{\parallel}(pred)|}{ \alpha_{\parallel}(mock)}
\end{equation}
\begin{equation}
d_{\alpha_\perp}=\frac{|\alpha_{\perp}(mock)-\alpha_{\perp}(pred)|}{ \alpha_{\perp}(mock)}
\end{equation}
Where $a_{\parallel}(mock)$ is the fitting results of {\color{black}$\alpha_\parallel$} %ap 
parameter by using the mock DLA catalog, and $a_{\parallel}(pred)$ stands for the results using our DLA catalog. 
Similarly, $a_{\perp}(mock)$ and $a_{\perp}(pred)$ are fitting value of {\color{black}$\alpha_\perp$} parameter using mock DLA catalog and our DLA catalog. 
{\color{black}
We also define the ratio of difference and error as below:
\begin{equation}
{ratio}_{\alpha_\parallel}=\frac{|\alpha_{\parallel}(mock)-\alpha_{\parallel}(pred)|}{\sigma \alpha_{\parallel}(mock)}
\end{equation}
\begin{equation}
{ratio}_{\alpha_\perp}=\frac{|\alpha_{\perp}(mock)-\alpha_{\perp}(pred)|}{\sigma \alpha_{\perp}(mock)}
\end{equation}
The differences between these two fitting results are less than $0.61\%$. This difference is lower than the statistical error using the DESI first year mock spectra (above $1.7\%$ shown in Table \ref{tab:bao parameter}). The statistical error will be reduced in the next 4 years DESI survey. With increasing S/N in the next few years, the DLA finder can give a better detection than the first year. And this difference can be further reduced. }The DLA catalog generated by our CNN model can be used for the BAO fitting analysis. 

\begin{figure*}
\plotone{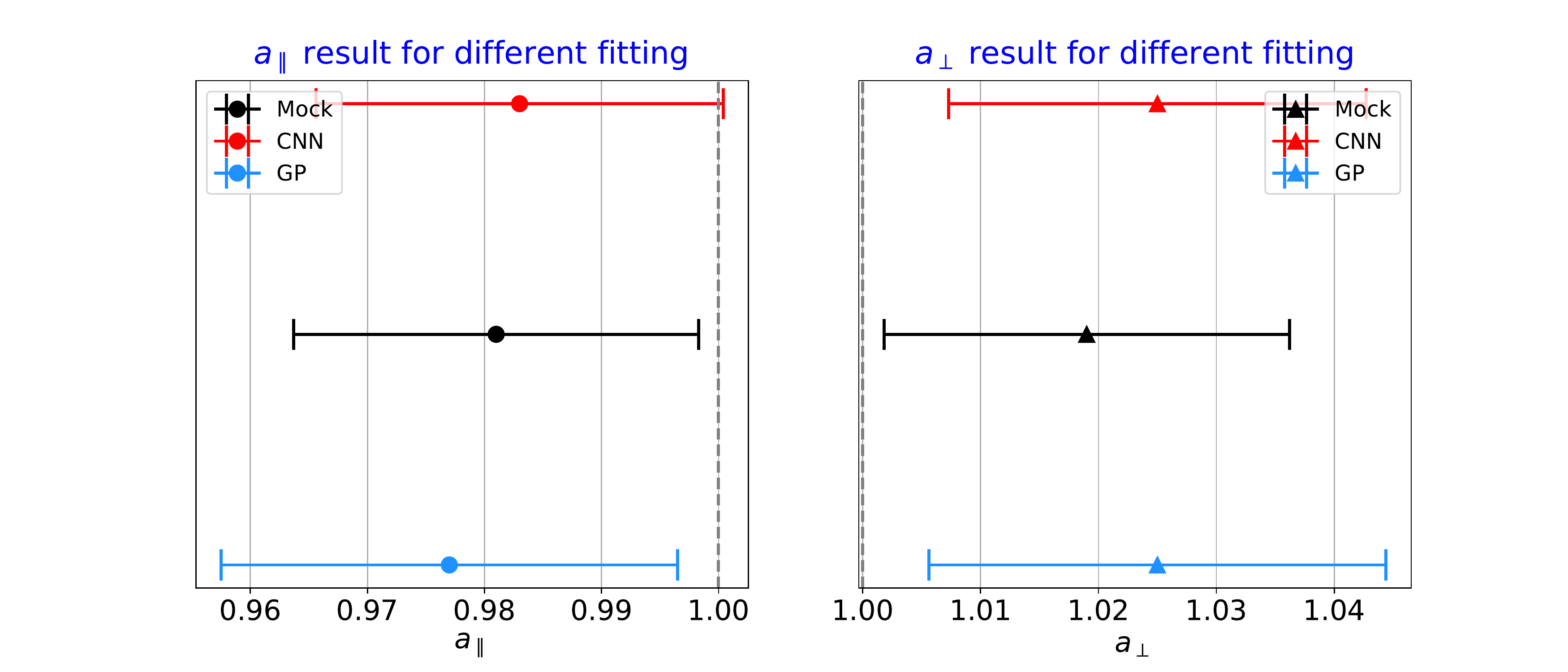}
\caption{The value and error bar for three different fittings: masking DLAs in the mock catalog, masking DLAs detected by CNN and masking DLAs detected by GP. It is clear that the result from masking GP DLA catalog has larger difference between the mock DLA catalog. The grey line is the theoretical value for these two parameters.   \label{fig:apaterrorbar}}
\end{figure*}

{\color{black}We also compare the performance on correlation fitting using DLA catalog generated by GP. The difference of best fitting parameters between GP and mock shown in Table \ref{tab:bao parameter} are about 0.25$\%$ larger than the difference between CNN and mock. But this difference is still less than the statistical error. So the BAO fitting result for the mock spectra is not affected much in any of the analyses presented. 
%in Table \ref{tab:bao parameter}. 
CNN DLA catalog can give a better BAO analysis support because of the higher completeness. The difference of fitting is clear when checking the difference of best fitting parameters value.  We have also plotted the result in Figure \ref{fig:apaterrorbar}. }{\color{black} Details about more parameters are shown in \ref{tab:appendix parameter}. 
%Taking all the parameters into account, the fitting result using CNN DLA catalog still performs smallest difference between the results using mock DLA catalog. 
This BAO analysis result is based on the mock spectra for DESI first year's survey. We plan to use different mock spectra to do the fitting comparison in the future. }

\section{Conclusion} \label{Conclusion}
In this manuscript, we have applied the deep learning techniques on QSO spectra to classify and characterize the damped Ly$\alpha$ systems (DLAs). We improved the CNN model created by \citet{2018MNRAS.476.1151P}
using DESI mock spectra to make it work successfully on DESI spectra. {\color{black}We optimize the preprocess, training procedure, parameter selection and the performance on low S/N and high column density DLAs. }Our model can give effective and accurate estimation about redshift and column density. We also improve the performance for CNN on low S/N spectra by smoothing the input flux,this method may also be used to other algorithm for low S/N signals. This CNN model can detect DLAs even when the S/N of DLAs is only about 1. {\color{black}We believe that there is still room to further improve this algorithm. }
%We believe that there is still some potential for this fully-automated algorithm.

Besides, our DLA catalog can also help to do the BAO analysis.  When we want to use correlation to do the BAO fitting, a DLA catalog is necessary. The results produced by our DLA catalog are very close to the real mock results, within the difference about $0.61\%$ for the best fitting parameters. 

\textcolor{black}{Finally, we compare our CNN DLA Finder with the Gaussian Process model from \citet{2020MNRAS.496.5436H} based on DESI mock spectra. Note that it may not be possible to show all the advantages and disadvantages of the two models due to the limitation of the mock. However, It is sufficient to see the differences between the two models by comparing their predictions with the mock truths.} {\color{black}CNN model has a higher completeness and purity on detecting DLAs. 
The fitting differences using CNN and GP DLA catalog are less than the statistical error on BAO fitting using Y1 mock spectra. The CNN results yield a systematic uncertainty $0.25\%$ less than that of GP. 
The BAO fitting result is not affected much by using either DLA finders. Both algorithms can estimate the redshift of DLAs well. }While the Gaussian Process model has more accurate column density estimation. Combining the catalogs given by the two models, we can obtain a credible DLA catalog that can be widely used for real DESI spectra release.
%Our DLA finder will be widely used when the real DESI spectra releases. 

\section{Ackowledgement}
This research is supported by the Director, Office of Science, Office of High Energy Physics of the U.S. Department of Energy under Contract No. DE–AC02–05CH11231, and by the National Energy Research Scientific Computing Center, a DOE Office of Science User Facility under the same contract; additional support for DESI is provided by the U.S. National Science Foundation, Division of Astronomical Sciences under Contract No. AST-0950945 to the NSF’s National Optical-Infrared Astronomy Research Laboratory; the Science and Technologies Facilities Council of the United Kingdom; the Gordon and Betty Moore Foundation; the Heising-Simons Foundation; the French Alternative Energies and Atomic Energy Commission (CEA); the National Council of Science and Technology of Mexico; the Ministry of Economy of Spain, and by the DESI Member Institutions. The authors are honored to be permitted to conduct scientific research on Iolkam Du’ag (Kitt Peak), a mountain with particular significance to the Tohono O’odham Nation. BW, JZ, ZC supported by the National Key R$\&$D Program of China
(grant No.2018YFA0404503), the National Science Foundation of China (grant No. 12073014). AFR acknowledges support FSE funds trough the program Ramon y Cajal (RYC-2018-025210) of the Spanish Ministry of Science and Innovation. VI is supported by the Kavli foundation. BW, JZ, ZC acknowledge the fruitful discussion with Dr. Tao Qin at Microsoft Research Asia. BW and JZ appreciate the great help from Huaizhe Xu in International Digital Economy Academy(IDEA). {\color{black}The authors also appreciate  Mr. Ming-Feng Ho from University of California Riverside for the great help about running Gaussian Process algorithm.  }

\appendix
\section{BAO combine fitting aprameters}

\begin{deluxetable}{ccccccc}
\tablecaption{\color{black}{Best fitting parameters \label{tab:appendix parameter}}}
\tablehead{
\colhead{parameters} & \colhead{DLA mock} &\colhead{DLA CNN} &\colhead{DLA GP} &\colhead{No Mask}
}
\startdata
$\alpha_\parallel$  &  0.981  & 0.983 & 0.977 & 0.992\\
$\sigma$               & 0.0173 & 0.0174 & 0.0195 & 0.0212\\
$\alpha_\perp$      &  1.019 & 1.025 & 1.025 & 1.028\\
$\sigma$               & 0.0172 & 0.0177 & 0.0194 & 0.0227\\
$beta_{LYA}$          & 1.5603  & 1.5515  & 1.6684  & 1.4066  \\
$\sigma$               & 0.0259 & 0.0265 & 0.0584 & 0.0214 \\
$bias_{LYA}$          & -0.1359  & -0.1391  & -0.0755  & -0.1581 \\
$\sigma$               & 0.0013 &  0.0014 &  0.0018 &  0.0015  \\
$bias_{eta}(LYA)$    & -0.2186  & -0.2224 & -0.2077  & -0.2292  \\
$\sigma$               & 0.0021 & 0.0022 & 0.0028 & 0.0021\\
$beta_{HCD}$          & 0.8237  & 0.8703   & 0.8710  & 0.8891  \\
$\sigma$              & 0.0689 & 0.0685 & 0.0850 & 0.0657\\
$bias_{HCD}$          & -0.0591  & -0.0621  & -0.0701  & -0.0795 \\
$\sigma$             & 0.0022 & 0.0022 & 0.0022 & 0.0028 \\
$\chi^{2}/DOF$      & 1.081 & 1.091 & 1.098 & 1.136\\
%\caption{Shown are the values and standard deviation for ap, at and the reduced \chi^{2} in two different fittings}
\enddata
%1.096 1.071 1.073 1.025
%1.211 1.191 1.191 1.136
\end{deluxetable}

\section{Purity and Completeness for desiY1-0.14 mock spectra}
\begin{figure*}
\gridline{\fig{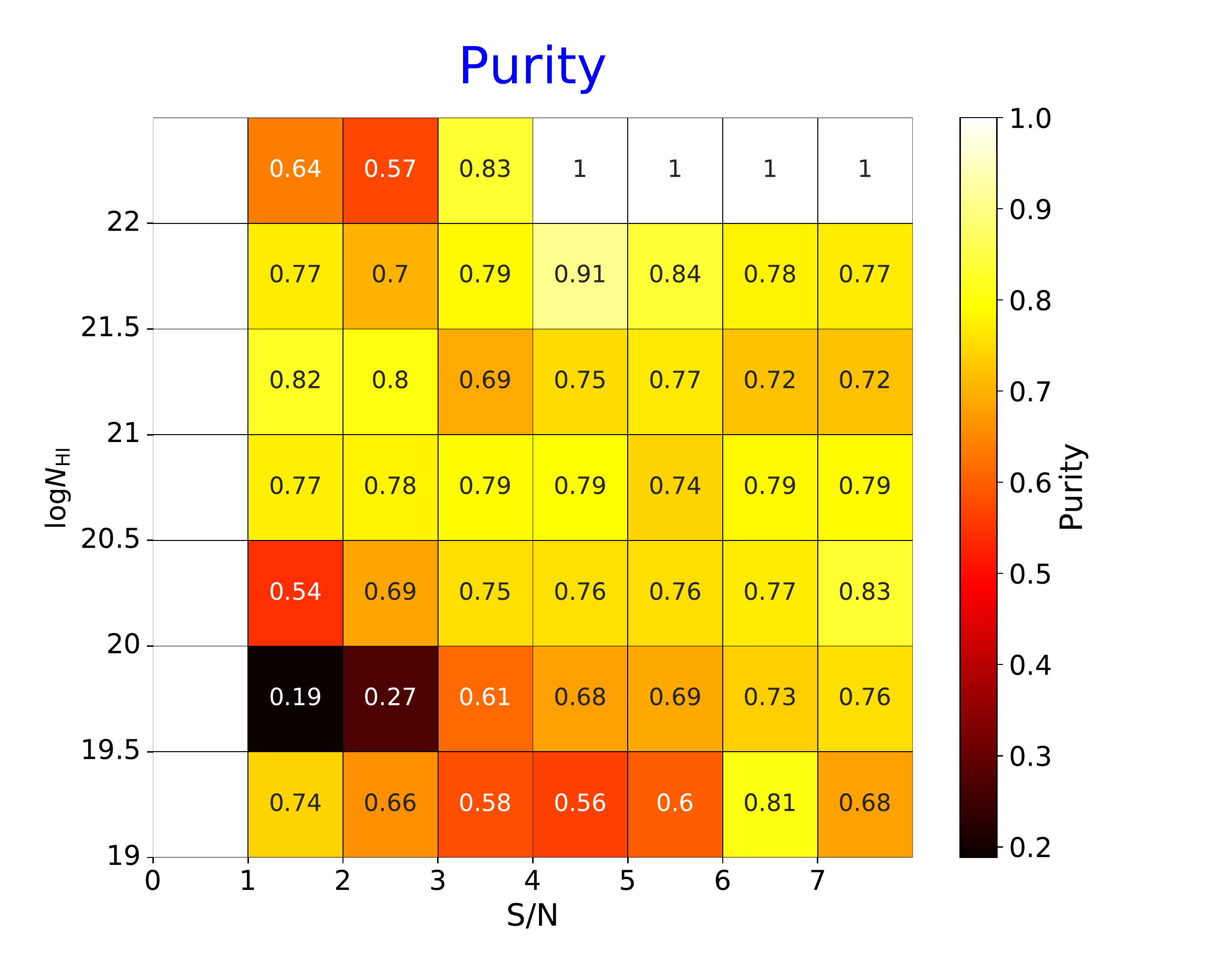}{0.5\textwidth}{(a)}
         \fig{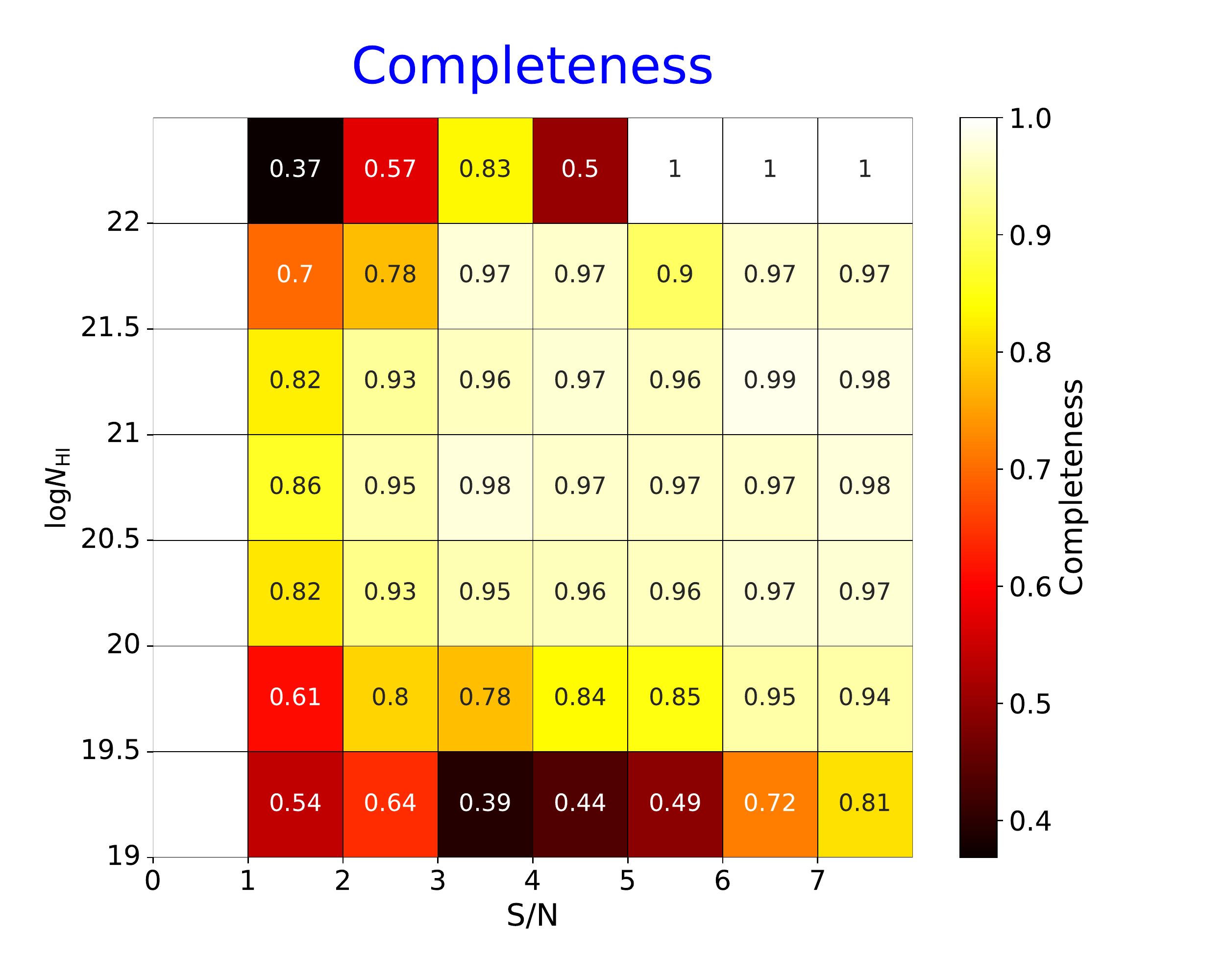}{0.5\textwidth}{(b)}
          }
\caption{We have applied our CNN model on desiY1-0.14 mock spectra. This mock spectra contains both DLAs and BALs. The purity and completeness are shown in (a) and (b). The minimum S/N for this mock is 1.0, and thus the left panel (S/N $<0$) is blank. The completeness is still above 96$\%$ for S/N$>$3 spectra. The purity drops about 10$\%$ to 20$\%$ in different bins.  Nevertheless, the DESI has a formal BAL catalog which will get rid of more than $98.6\%$ BALs from the catalog. Then, we can run DLA finder on the BAL-removed spectra. Therefore, we think that the purity result shown in Figure\ref{fig:purity} is still valid.    \label{fig:bal-pc}}
\end{figure*}

\section{DLA detection in SDSS spectra}
\begin{figure}
\gridline{\fig{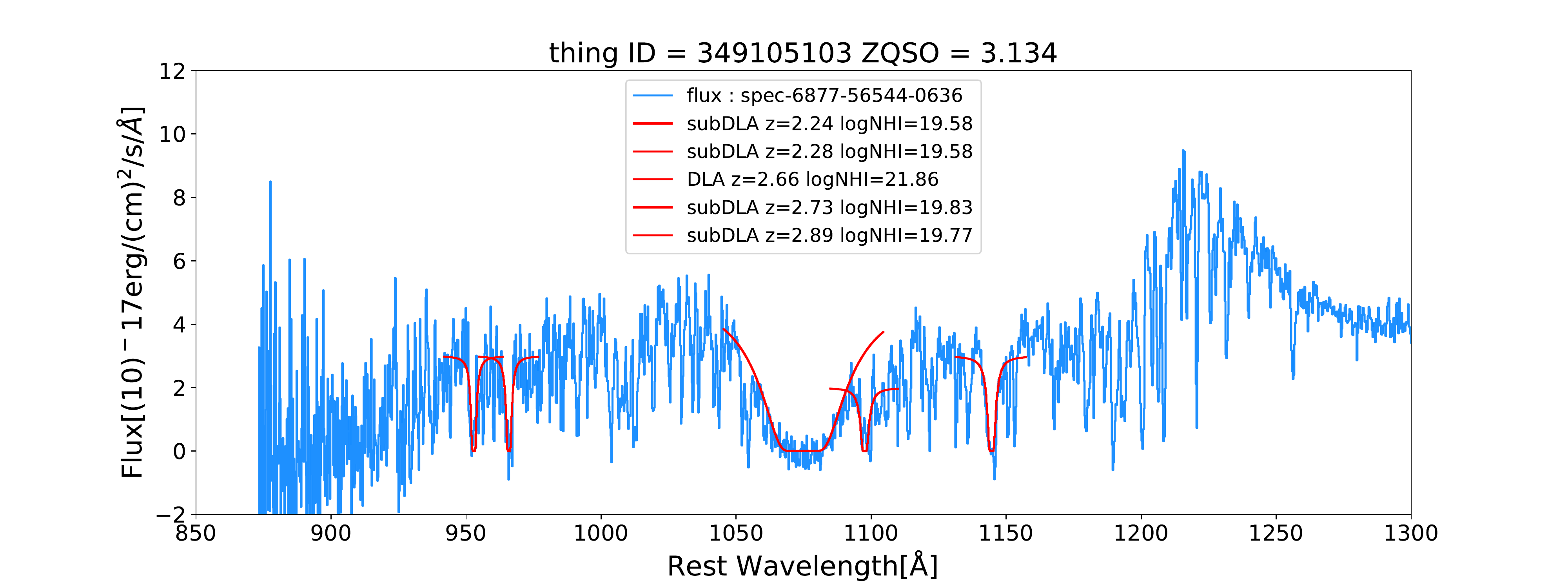}{0.9\textwidth}{(a)}
          }
\gridline{\fig{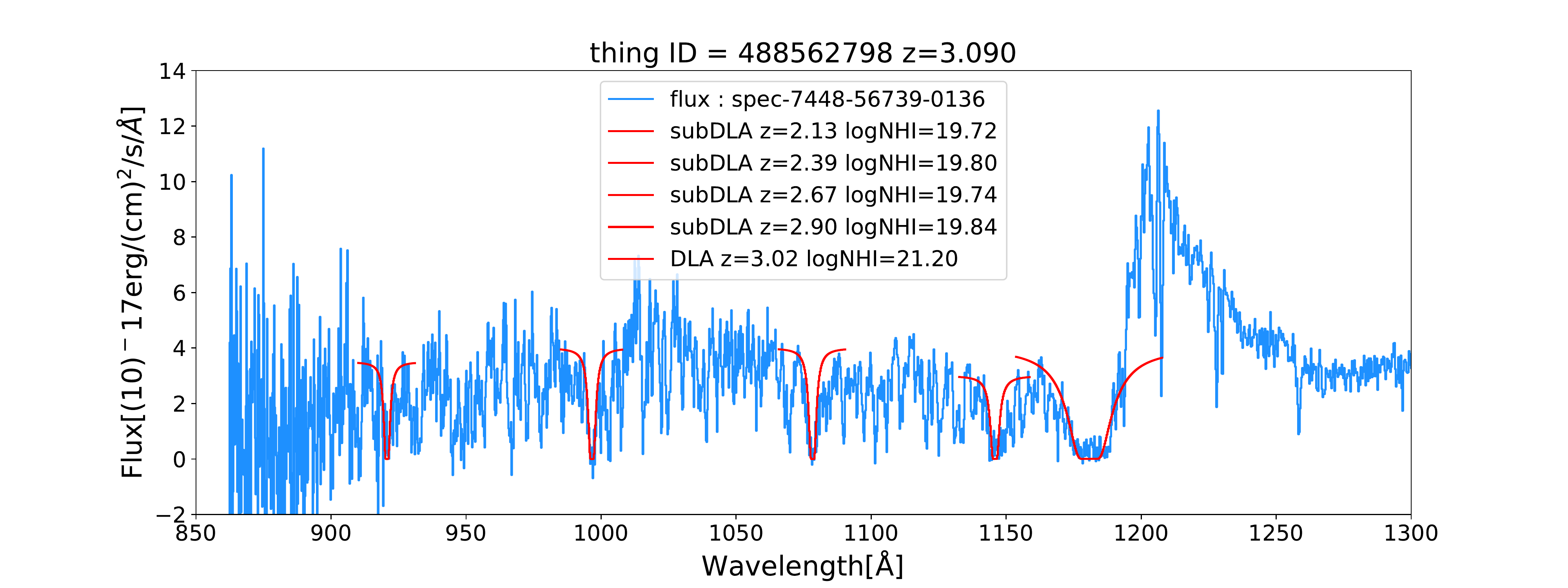}{0.9\textwidth}{(b)}
          }
\caption{We have applied our CNN model on these two spectra mentioned by \citet{2021arXiv210310964H}. It is clear that the DLA and subDLA candidates are detected.   \label{fig:sdss}}
\end{figure}

In \citet{2021arXiv210310964H}, the author mentioned that the previous version of CNN DLA finder\citet{2018MNRAS.476.1151P} missed high column density DLAs in two SDSS spectra. We test our model on these two spectra, both the DLAs and subDLAs can be detected as shown in Figure \ref{fig:sdss}

\bibliography{sample631}{}

\begin{thebibliography}{}
\expandafter\ifx\csname natexlab\endcsname\relax\def\natexlab#1{#1}\fi
\providecommand{\url}[1]{\href{#1}{#1}}
\providecommand{\dodoi}[1]{doi:~\href{http://doi.org/#1}{\nolinkurl{#1}}}
\providecommand{\doeprint}[1]{\href{http://ascl.net/#1}{\nolinkurl{http://ascl.net/#1}}}
\providecommand{\doarXiv}[1]{\href{https://arxiv.org/abs/#1}{\nolinkurl{https://arxiv.org/abs/#1}}}

\bibitem[{{Bird} {et~al.}(2014){Bird}, {Vogelsberger}, {Haehnelt}, {Sijacki},
  {Genel}, {Torrey}, {Springel}, \& {Hernquist}}]{2014MNRAS.445.2313B}
{Bird}, S., {Vogelsberger}, M., {Haehnelt}, M., {et~al.} 2014, \mnras, 445,
  2313, \dodoi{10.1093/mnras/stu1923}

\bibitem[{{Busca} {et~al.}(2013){Busca}, {Delubac}, {Rich}, {Bailey},
  {Font-Ribera}, {Kirkby}, {Le Goff}, {Pieri}, {Slosar}, {Aubourg}, {Bautista},
  {Bizyaev}, {Blomqvist}, {Bolton}, {Bovy}, {Brewington}, {Borde}, {Brinkmann},
  {Carithers}, {Croft}, {Dawson}, {Ebelke}, {Eisenstein}, {Hamilton}, {Ho},
  {Hogg}, {Honscheid}, {Lee}, {Lundgren}, {Malanushenko}, {Malanushenko},
  {Margala}, {Maraston}, {Mehta}, {Miralda-Escud{\'e}}, {Myers}, {Nichol},
  {Noterdaeme}, {Olmstead}, {Oravetz}, {Palanque-Delabrouille}, {Pan},
  {P{\^a}ris}, {Percival}, {Petitjean}, {Roe}, {Rollinde}, {Ross}, {Rossi},
  {Schlegel}, {Schneider}, {Shelden}, {Sheldon}, {Simmons}, {Snedden},
  {Tinker}, {Viel}, {Weaver}, {Weinberg}, {White}, {Y{\`e}che}, \&
  {York}}]{2013A&A...552A..96B}
{Busca}, N.~G., {Delubac}, T., {Rich}, J., {et~al.} 2013, \aap, 552, A96,
  \dodoi{10.1051/0004-6361/201220724}

\bibitem[{{Cai} {et~al.}(2016){Cai}, {Fan}, {Peirani}, {Bian}, {Frye},
  {McGreer}, {Prochaska}, {Lau}, {Tejos}, {Ho}, \&
  {Schneider}}]{2016ApJ...833..135C}
{Cai}, Z., {Fan}, X., {Peirani}, S., {et~al.} 2016, \apj, 833, 135,
  \dodoi{10.3847/1538-4357/833/2/135}

\bibitem[{{Cai} {et~al.}(2017){Cai}, {Fan}, {Bian}, {Zabludoff}, {Yang},
  {Prochaska}, {McGreer}, {Zheng}, {Kashikawa}, {Wang}, {Frye}, {Green}, \&
  {Jiang}}]{2017ApJ...839..131C}
{Cai}, Z., {Fan}, X., {Bian}, F., {et~al.} 2017, \apj, 839, 131,
  \dodoi{10.3847/1538-4357/aa6a1a}

\bibitem[{{Chabanier} {et~al.}(2021){Chabanier}, {Etourneau}, {Le Goff},
  {Rich}, {Stermer}, {Abolfathi}, {Font-Ribera}, {Gonzalez-Morales}, {de la
  Macorra}, {P{\'e}rez-R{\'a}fols}, {Petitjean}, {Pieri}, {Ravoux}, {Rossi}, \&
  {Schneider}}]{2021arXiv210709612C}
{Chabanier}, S., {Etourneau}, T., {Le Goff}, J.-M., {et~al.} 2021, arXiv
  e-prints, arXiv:2107.09612.
\newblock \doarXiv{2107.09612}

\bibitem[{{DESI Collaboration} {et~al.}(2016{\natexlab{a}}){DESI
  Collaboration}, {Aghamousa}, {Aguilar}, {Ahlen}, {Alam}, {Allen}, {Allende
  Prieto}, {Annis}, {Bailey}, {Balland}, {Ballester}, {Baltay}, {Beaufore},
  {Bebek}, {Beers}, {Bell}, {Bernal}, {Besuner}, {Beutler}, {Blake}, {Bleuler},
  {Blomqvist}, {Blum}, {Bolton}, {Briceno}, {Brooks}, {Brownstein},
  {Buckley-Geer}, {Burden}, {Burtin}, {Busca}, {Cahn}, {Cai}, {Cardiel-Sas},
  {Carlberg}, {Carton}, {Casas}, {Castander}, {Cervantes-Cota}, {Claybaugh},
  {Close}, {Coker}, {Cole}, {Comparat}, {Cooper}, {Cousinou}, {Crocce}, {Cuby},
  {Cunningham}, {Davis}, {Dawson}, {de la Macorra}, {De Vicente}, {Delubac},
  {Derwent}, {Dey}, {Dhungana}, {Ding}, {Doel}, {Duan}, {Ealet}, {Edelstein},
  {Eftekharzadeh}, {Eisenstein}, {Elliott}, {Escoffier}, {Evatt}, {Fagrelius},
  {Fan}, {Fanning}, {Farahi}, {Farihi}, {Favole}, {Feng}, {Fernandez},
  {Findlay}, {Finkbeiner}, {Fitzpatrick}, {Flaugher}, {Flender}, {Font-Ribera},
  {Forero-Romero}, {Fosalba}, {Frenk}, {Fumagalli}, {Gaensicke}, {Gallo},
  {Garcia-Bellido}, {Gaztanaga}, {Pietro Gentile Fusillo}, {Gerard},
  {Gershkovich}, {Giannantonio}, {Gillet}, {Gonzalez-de-Rivera},
  {Gonzalez-Perez}, {Gott}, {Graur}, {Gutierrez}, {Guy}, {Habib}, {Heetderks},
  {Heetderks}, {Heitmann}, {Hellwing}, {Herrera}, {Ho}, {Holland}, {Honscheid},
  {Huff}, {Hutchinson}, {Huterer}, {Hwang}, {Illa Laguna}, {Ishikawa},
  {Jacobs}, {Jeffrey}, {Jelinsky}, {Jennings}, {Jiang}, {Jimenez}, {Johnson},
  {Joyce}, {Jullo}, {Juneau}, {Kama}, {Karcher}, {Karkar}, {Kehoe}, {Kennamer},
  {Kent}, {Kilbinger}, {Kim}, {Kirkby}, {Kisner}, {Kitanidis}, {Kneib},
  {Koposov}, {Kovacs}, {Koyama}, {Kremin}, {Kron}, {Kronig}, {Kueter-Young},
  {Lacey}, {Lafever}, {Lahav}, {Lambert}, {Lampton}, {Landriau}, {Lang},
  {Lauer}, {Le Goff}, {Le Guillou}, {Le Van Suu}, {Lee}, {Lee}, {Leitner},
  {Lesser}, {Levi}, {L'Huillier}, {Li}, {Liang}, {Lin}, {Linder}, {Loebman},
  {Luki{\'c}}, {Ma}, {MacCrann}, {Magneville}, {Makarem}, {Manera}, {Manser},
  {Marshall}, {Martini}, {Massey}, {Matheson}, {McCauley}, {McDonald},
  {McGreer}, {Meisner}, {Metcalfe}, {Miller}, {Miquel}, {Moustakas}, {Myers},
  {Naik}, {Newman}, {Nichol}, {Nicola}, {Nicolati da Costa}, {Nie}, {Niz},
  {Norberg}, {Nord}, {Norman}, {Nugent}, {O'Brien}, {Oh}, {Olsen}, {Padilla},
  {Padmanabhan}, {Padmanabhan}, {Palanque-Delabrouille}, {Palmese},
  {Pappalardo}, {P{\^a}ris}, {Park}, {Patej}, {Peacock}, {Peiris}, {Peng},
  {Percival}, {Perruchot}, {Pieri}, {Pogge}, {Pollack}, {Poppett}, {Prada},
  {Prakash}, {Probst}, {Rabinowitz}, {Raichoor}, {Ree}, {Refregier}, {Regal},
  {Reid}, {Reil}, {Rezaie}, {Rockosi}, {Roe}, {Ronayette}, {Roodman}, {Ross},
  {Ross}, {Rossi}, {Rozo}, {Ruhlmann-Kleider}, {Rykoff}, {Sabiu}, {Samushia},
  {Sanchez}, {Sanchez}, {Schlegel}, {Schneider}, {Schubnell}, {Secroun},
  {Seljak}, {Seo}, {Serrano}, {Shafieloo}, {Shan}, {Sharples}, {Sholl},
  {Shourt}, {Silber}, {Silva}, {Sirk}, {Slosar}, {Smith}, {Smoot}, {Som},
  {Song}, {Sprayberry}, {Staten}, {Stefanik}, {Tarle}, {Sien Tie}, {Tinker},
  {Tojeiro}, {Valdes}, {Valenzuela}, {Valluri}, {Vargas-Magana}, {Verde},
  {Walker}, {Wang}, {Wang}, {Weaver}, {Weaverdyck}, {Wechsler}, {Weinberg},
  {White}, {Yang}, {Yeche}, {Zhang}, {Zhao}, {Zheng}, {Zhou}, {Zhou}, {Zhu},
  {Zou}, \& {Zu}}]{2016arXiv161100036D}
{DESI Collaboration}, {Aghamousa}, A., {Aguilar}, J., {et~al.}
  2016{\natexlab{a}}, arXiv e-prints, arXiv:1611.00036.
\newblock \doarXiv{1611.00036}

\bibitem[{{DESI Collaboration} {et~al.}(2016{\natexlab{b}}){DESI
  Collaboration}, {Aghamousa}, {Aguilar}, {Ahlen}, {Alam}, {Allen}, {Allende
  Prieto}, {Annis}, {Bailey}, {Balland}, {Ballester}, {Baltay}, {Beaufore},
  {Bebek}, {Beers}, {Bell}, {Bernal}, {Besuner}, {Beutler}, {Blake}, {Bleuler},
  {Blomqvist}, {Blum}, {Bolton}, {Briceno}, {Brooks}, {Brownstein},
  {Buckley-Geer}, {Burden}, {Burtin}, {Busca}, {Cahn}, {Cai}, {Cardiel-Sas},
  {Carlberg}, {Carton}, {Casas}, {Castander}, {Cervantes-Cota}, {Claybaugh},
  {Close}, {Coker}, {Cole}, {Comparat}, {Cooper}, {Cousinou}, {Crocce}, {Cuby},
  {Cunningham}, {Davis}, {Dawson}, {de la Macorra}, {De Vicente}, {Delubac},
  {Derwent}, {Dey}, {Dhungana}, {Ding}, {Doel}, {Duan}, {Ealet}, {Edelstein},
  {Eftekharzadeh}, {Eisenstein}, {Elliott}, {Escoffier}, {Evatt}, {Fagrelius},
  {Fan}, {Fanning}, {Farahi}, {Farihi}, {Favole}, {Feng}, {Fernandez},
  {Findlay}, {Finkbeiner}, {Fitzpatrick}, {Flaugher}, {Flender}, {Font-Ribera},
  {Forero-Romero}, {Fosalba}, {Frenk}, {Fumagalli}, {Gaensicke}, {Gallo},
  {Garcia-Bellido}, {Gaztanaga}, {Pietro Gentile Fusillo}, {Gerard},
  {Gershkovich}, {Giannantonio}, {Gillet}, {Gonzalez-de-Rivera},
  {Gonzalez-Perez}, {Gott}, {Graur}, {Gutierrez}, {Guy}, {Habib}, {Heetderks},
  {Heetderks}, {Heitmann}, {Hellwing}, {Herrera}, {Ho}, {Holland}, {Honscheid},
  {Huff}, {Hutchinson}, {Huterer}, {Hwang}, {Illa Laguna}, {Ishikawa},
  {Jacobs}, {Jeffrey}, {Jelinsky}, {Jennings}, {Jiang}, {Jimenez}, {Johnson},
  {Joyce}, {Jullo}, {Juneau}, {Kama}, {Karcher}, {Karkar}, {Kehoe}, {Kennamer},
  {Kent}, {Kilbinger}, {Kim}, {Kirkby}, {Kisner}, {Kitanidis}, {Kneib},
  {Koposov}, {Kovacs}, {Koyama}, {Kremin}, {Kron}, {Kronig}, {Kueter-Young},
  {Lacey}, {Lafever}, {Lahav}, {Lambert}, {Lampton}, {Landriau}, {Lang},
  {Lauer}, {Le Goff}, {Le Guillou}, {Le Van Suu}, {Lee}, {Lee}, {Leitner},
  {Lesser}, {Levi}, {L'Huillier}, {Li}, {Liang}, {Lin}, {Linder}, {Loebman},
  {Luki{\'c}}, {Ma}, {MacCrann}, {Magneville}, {Makarem}, {Manera}, {Manser},
  {Marshall}, {Martini}, {Massey}, {Matheson}, {McCauley}, {McDonald},
  {McGreer}, {Meisner}, {Metcalfe}, {Miller}, {Miquel}, {Moustakas}, {Myers},
  {Naik}, {Newman}, {Nichol}, {Nicola}, {Nicolati da Costa}, {Nie}, {Niz},
  {Norberg}, {Nord}, {Norman}, {Nugent}, {O'Brien}, {Oh}, {Olsen}, {Padilla},
  {Padmanabhan}, {Padmanabhan}, {Palanque-Delabrouille}, {Palmese},
  {Pappalardo}, {P{\^a}ris}, {Park}, {Patej}, {Peacock}, {Peiris}, {Peng},
  {Percival}, {Perruchot}, {Pieri}, {Pogge}, {Pollack}, {Poppett}, {Prada},
  {Prakash}, {Probst}, {Rabinowitz}, {Raichoor}, {Ree}, {Refregier}, {Regal},
  {Reid}, {Reil}, {Rezaie}, {Rockosi}, {Roe}, {Ronayette}, {Roodman}, {Ross},
  {Ross}, {Rossi}, {Rozo}, {Ruhlmann-Kleider}, {Rykoff}, {Sabiu}, {Samushia},
  {Sanchez}, {Sanchez}, {Schlegel}, {Schneider}, {Schubnell}, {Secroun},
  {Seljak}, {Seo}, {Serrano}, {Shafieloo}, {Shan}, {Sharples}, {Sholl},
  {Shourt}, {Silber}, {Silva}, {Sirk}, {Slosar}, {Smith}, {Smoot}, {Som},
  {Song}, {Sprayberry}, {Staten}, {Stefanik}, {Tarle}, {Sien Tie}, {Tinker},
  {Tojeiro}, {Valdes}, {Valenzuela}, {Valluri}, {Vargas-Magana}, {Verde},
  {Walker}, {Wang}, {Wang}, {Weaver}, {Weaverdyck}, {Wechsler}, {Weinberg},
  {White}, {Yang}, {Yeche}, {Zhang}, {Zhao}, {Zheng}, {Zhou}, {Zhou}, {Zhu},
  {Zou}, \& {Zu}}]{2016arXiv161100037D}
---. 2016{\natexlab{b}}, arXiv e-prints, arXiv:1611.00037.
\newblock \doarXiv{1611.00037}

\bibitem[{{Dey} {et~al.}(2019){Dey}, {Schlegel}, {Lang}, {Blum}, {Burleigh},
  {Fan}, {Findlay}, {Finkbeiner}, {Herrera}, {Juneau}, {Landriau}, {Levi},
  {McGreer}, {Meisner}, {Myers}, {Moustakas}, {Nugent}, {Patej}, {Schlafly},
  {Walker}, {Valdes}, {Weaver}, {Y{\`e}che}, {Zou}, {Zhou}, {Abareshi},
  {Abbott}, {Abolfathi}, {Aguilera}, {Alam}, {Allen}, {Alvarez}, {Annis},
  {Ansarinejad}, {Aubert}, {Beechert}, {Bell}, {BenZvi}, {Beutler}, {Bielby},
  {Bolton}, {Brice{\~n}o}, {Buckley-Geer}, {Butler}, {Calamida}, {Carlberg},
  {Carter}, {Casas}, {Castander}, {Choi}, {Comparat}, {Cukanovaite}, {Delubac},
  {DeVries}, {Dey}, {Dhungana}, {Dickinson}, {Ding}, {Donaldson}, {Duan},
  {Duckworth}, {Eftekharzadeh}, {Eisenstein}, {Etourneau}, {Fagrelius},
  {Farihi}, {Fitzpatrick}, {Font-Ribera}, {Fulmer}, {G{\"a}nsicke},
  {Gaztanaga}, {George}, {Gerdes}, {Gontcho}, {Gorgoni}, {Green}, {Guy},
  {Harmer}, {Hernandez}, {Honscheid}, {Huang}, {James}, {Jannuzi}, {Jiang},
  {Joyce}, {Karcher}, {Karkar}, {Kehoe}, {Kneib}, {Kueter-Young}, {Lan},
  {Lauer}, {Le Guillou}, {Le Van Suu}, {Lee}, {Lesser}, {Perreault Levasseur},
  {Li}, {Mann}, {Marshall}, {Mart{\'\i}nez-V{\'a}zquez}, {Martini}, {du Mas des
  Bourboux}, {McManus}, {Meier}, {M{\'e}nard}, {Metcalfe},
  {Mu{\~n}oz-Guti{\'e}rrez}, {Najita}, {Napier}, {Narayan}, {Newman}, {Nie},
  {Nord}, {Norman}, {Olsen}, {Paat}, {Palanque-Delabrouille}, {Peng},
  {Poppett}, {Poremba}, {Prakash}, {Rabinowitz}, {Raichoor}, {Rezaie},
  {Robertson}, {Roe}, {Ross}, {Ross}, {Rudnick}, {Safonova}, {Saha},
  {S{\'a}nchez}, {Savary}, {Schweiker}, {Scott}, {Seo}, {Shan}, {Silva},
  {Slepian}, {Soto}, {Sprayberry}, {Staten}, {Stillman}, {Stupak}, {Summers},
  {Sien Tie}, {Tirado}, {Vargas-Maga{\~n}a}, {Vivas}, {Wechsler}, {Williams},
  {Yang}, {Yang}, {Yapici}, {Zaritsky}, {Zenteno}, {Zhang}, {Zhang}, {Zhou}, \&
  {Zhou}}]{2019AJ....157..168D}
{Dey}, A., {Schlegel}, D.~J., {Lang}, D., {et~al.} 2019, \aj, 157, 168,
  \dodoi{10.3847/1538-3881/ab089d}

\bibitem[{{Draine}(2011)}]{2011piim.book.....D}
{Draine}, B.~T. 2011, {Physics of the Interstellar and Intergalactic Medium}

\bibitem[{{du Mas des Bourboux} {et~al.}(2020){du Mas des Bourboux}, {Rich},
  {Font-Ribera}, {de Sainte Agathe}, {Farr}, {Etourneau}, {Le Goff}, {Cuceu},
  {Balland}, {Bautista}, {Blomqvist}, {Brinkmann}, {Brownstein}, {Chabanier},
  {Chaussidon}, {Dawson}, {Gonz{\'a}lez-Morales}, {Guy}, {Lyke}, {de la
  Macorra}, {Mueller}, {Myers}, {Nitschelm}, {Mu{\~n}oz Guti{\'e}rrez},
  {Palanque-Delabrouille}, {Parker}, {Percival}, {P{\'e}rez-R{\`a}fols},
  {Petitjean}, {Pieri}, {Ravoux}, {Rossi}, {Schneider}, {Seo}, {Slosar},
  {Stermer}, {Vivek}, {Y{\`e}che}, \& {Youles}}]{2020ApJ...901..153D}
{du Mas des Bourboux}, H., {Rich}, J., {Font-Ribera}, A., {et~al.} 2020, \apj,
  901, 153, \dodoi{10.3847/1538-4357/abb085}

\bibitem[{{Farr} {et~al.}(2020){Farr}, {Font-Ribera}, {du Mas des Bourboux},
  {Mu{\~n}oz-Guti{\'e}rrez}, {S{\'a}nchez}, {Pontzen}, {Xochitl
  Gonz{\'a}lez-Morales}, {Alonso}, {Brooks}, {Doel}, {Etourneau}, {Guy}, {Le
  Goff}, {de la Macorra}, {Palanque-Delabrouille}, {P{\'e}rez-R{\`a}fols},
  {Rich}, {Slosar}, {Tarle}, {Yutong}, \& {Zhang}}]{2020JCAP...03..068F}
{Farr}, J., {Font-Ribera}, A., {du Mas des Bourboux}, H., {et~al.} 2020, \jcap,
  2020, 068, \dodoi{10.1088/1475-7516/2020/03/068}

\bibitem[{{Finley} {et~al.}(2013){Finley}, {Petitjean}, {P{\^a}ris},
  {Noterdaeme}, {Brinkmann}, {Myers}, {Ross}, {Schneider}, {Bizyaev},
  {Brewington}, {Ebelke}, {Malanushenko}, {Malanushenko}, {Oravetz}, {Pan},
  {Simmons}, \& {Snedden}}]{2013A&A...558A.111F}
{Finley}, H., {Petitjean}, P., {P{\^a}ris}, I., {et~al.} 2013, \aap, 558, A111,
  \dodoi{10.1051/0004-6361/201321745}

\bibitem[{{Font-Ribera} \& {Miralda-Escud{\'e}}(2012)}]{2012JCAP...07..028F}
{Font-Ribera}, A., \& {Miralda-Escud{\'e}}, J. 2012, \jcap, 2012, 028,
  \dodoi{10.1088/1475-7516/2012/07/028}

\bibitem[{{Fumagalli} {et~al.}(2011){Fumagalli}, {Prochaska}, {Kasen}, {Dekel},
  {Ceverino}, \& {Primack}}]{2011MNRAS.418.1796F}
{Fumagalli}, M., {Prochaska}, J.~X., {Kasen}, D., {et~al.} 2011, \mnras, 418,
  1796, \dodoi{10.1111/j.1365-2966.2011.19599.x}

\bibitem[{{Gardner} {et~al.}(1997){Gardner}, {Sharples}, {Frenk}, \&
  {Carrasco}}]{1997ApJ...480L..99G}
{Gardner}, J.~P., {Sharples}, R.~M., {Frenk}, C.~S., \& {Carrasco}, B.~E. 1997,
  \apjl, 480, L99, \dodoi{10.1086/310630}

\bibitem[{{Garnett} {et~al.}(2017){Garnett}, {Ho}, {Bird}, \&
  {Schneider}}]{2017MNRAS.472.1850G}
{Garnett}, R., {Ho}, S., {Bird}, S., \& {Schneider}, J. 2017, \mnras, 472,
  1850, \dodoi{10.1093/mnras/stx1958}

\bibitem[{{Gonzalez-Morales} \& {DESI Lyman $\alpha$ Working
  Group}(\ndd)}]{Gonzalez-Morales2021}
{Gonzalez-Morales}, A.~X., \& {DESI Lyman $\alpha$ Working Group}. \ndd

\bibitem[{{Grudi{\'c}} {et~al.}(2020){Grudi{\'c}}, {Guszejnov}, {Hopkins},
  {Offner}, \& {Faucher-Gigu{\`e}re}}]{2020arXiv201011254G}
{Grudi{\'c}}, M.~Y., {Guszejnov}, D., {Hopkins}, P.~F., {Offner}, S. S.~R., \&
  {Faucher-Gigu{\`e}re}, C.-A. 2020, arXiv e-prints, arXiv:2010.11254.
\newblock \doarXiv{2010.11254}

\bibitem[{{Guo} \& {Martini}(2019)}]{2019ApJ...879...72G}
{Guo}, Z., \& {Martini}, P. 2019, \apj, 879, 72,
  \dodoi{10.3847/1538-4357/ab2590}

\bibitem[{Herrera-Alcantar(2020)}]{Herrera-Alcantar2020}
Herrera-Alcantar, H.~K. 2020, Master's thesis, Universidad de Guanajuato

\bibitem[{{Ho} {et~al.}(2020){Ho}, {Bird}, \& {Garnett}}]{2020MNRAS.496.5436H}
{Ho}, M.-F., {Bird}, S., \& {Garnett}, R. 2020, \mnras, 496, 5436,
  \dodoi{10.1093/mnras/staa1806}

\bibitem[{{Ho} {et~al.}(2021){Ho}, {Bird}, \& {Garnett}}]{2021arXiv210310964H}
---. 2021, arXiv e-prints, arXiv:2103.10964.
\newblock \doarXiv{2103.10964}

\bibitem[{{Jolly} {et~al.}(2020){Jolly}, {Knudsen}, \&
  {Stanley}}]{2020MNRAS.499.3992J}
{Jolly}, J.-B., {Knudsen}, K.~K., \& {Stanley}, F. 2020, \mnras, 499, 3992,
  \dodoi{10.1093/mnras/staa2908}

\bibitem[{{Krogager} {et~al.}(2020){Krogager}, {M{\o}ller}, {Christensen},
  {Noterdaeme}, {Fynbo}, \& {Freudling}}]{2020MNRAS.495.3014K}
{Krogager}, J.-K., {M{\o}ller}, P., {Christensen}, L.~B., {et~al.} 2020,
  \mnras, 495, 3014, \dodoi{10.1093/mnras/staa1414}

\bibitem[{{Lee} {et~al.}(2020){Lee}, {Webb}, \&
  {Carswell}}]{2020MNRAS.491.5555L}
{Lee}, C.~C., {Webb}, J.~K., \& {Carswell}, R.~F. 2020, \mnras, 491, 5555,
  \dodoi{10.1093/mnras/stz3170}

\bibitem[{{Lee} {et~al.}(2014){Lee}, {Hennawi}, {Stark}, {Prochaska}, {White},
  {Schlegel}, {Eilers}, {Arinyo-i-Prats}, {Suzuki}, {Croft}, {Caputi},
  {Cassata}, {Ilbert}, {Garilli}, {Koekemoer}, {Le Brun}, {Le F{\`e}vre},
  {Maccagni}, {Nugent}, {Taniguchi}, {Tasca}, {Tresse}, {Zamorani}, \&
  {Zucca}}]{2014ApJ...795L..12L}
{Lee}, K.-G., {Hennawi}, J.~F., {Stark}, C., {et~al.} 2014, \apjl, 795, L12,
  \dodoi{10.1088/2041-8205/795/1/L12}

\bibitem[{{Li} {et~al.}(2021){Li}, {Horowitz}, \& {Cai}}]{2021ApJ...916...20L}
{Li}, Z., {Horowitz}, B., \& {Cai}, Z. 2021, \apj, 916, 20,
  \dodoi{10.3847/1538-4357/ac044a}

\bibitem[{{Liske} {et~al.}(1998){Liske}, {Webb}, \&
  {Carswell}}]{1998MNRAS.301..787L}
{Liske}, J., {Webb}, J.~K., \& {Carswell}, R.~F. 1998, \mnras, 301, 787,
  \dodoi{10.1046/j.1365-8711.1998.02048.x}

\bibitem[{{McDonald}(2003)}]{2003ApJ...585...34M}
{McDonald}, P. 2003, \apj, 585, 34, \dodoi{10.1086/345945}

\bibitem[{{McGreer} {et~al.}(2013){McGreer}, {Jiang}, {Fan}, {Richards},
  {Strauss}, {Ross}, {White}, {Shen}, {Schneider}, {Myers}, {Brandt}, {DeGraf},
  {Glikman}, {Ge}, \& {Streblyanska}}]{2013ApJ...768..105M}
{McGreer}, I.~D., {Jiang}, L., {Fan}, X., {et~al.} 2013, \apj, 768, 105,
  \dodoi{10.1088/0004-637X/768/2/105}

\bibitem[{{McQuinn}(2016)}]{2016ARA&A..54..313M}
{McQuinn}, M. 2016, \araa, 54, 313, \dodoi{10.1146/annurev-astro-082214-122355}

\bibitem[{{Noterdaeme} {et~al.}(2019){Noterdaeme}, {Balashev}, {Krogager},
  {Srianand}, {Fathivavsari}, {Petitjean}, \& {Ledoux}}]{2019A&A...627A..32N}
{Noterdaeme}, P., {Balashev}, S., {Krogager}, J.~K., {et~al.} 2019, \aap, 627,
  A32, \dodoi{10.1051/0004-6361/201935371}

\bibitem[{{Noterdaeme} {et~al.}(2009){Noterdaeme}, {Petitjean}, {Ledoux}, \&
  {Srianand}}]{2009A&A...505.1087N}
{Noterdaeme}, P., {Petitjean}, P., {Ledoux}, C., \& {Srianand}, R. 2009, \aap,
  505, 1087, \dodoi{10.1051/0004-6361/200912768}

\bibitem[{{Noterdaeme} {et~al.}(2012{\natexlab{a}}){Noterdaeme}, {Petitjean},
  {Carithers}, {P{\^a}ris}, {Font-Ribera}, {Bailey}, {Aubourg}, {Bizyaev},
  {Ebelke}, {Finley}, {Ge}, {Malanushenko}, {Malanushenko},
  {Miralda-Escud{\'e}}, {Myers}, {Oravetz}, {Pan}, {Pieri}, {Ross},
  {Schneider}, {Simmons}, \& {York}}]{2012A&A...547L...1N}
{Noterdaeme}, P., {Petitjean}, P., {Carithers}, W.~C., {et~al.}
  2012{\natexlab{a}}, \aap, 547, L1, \dodoi{10.1051/0004-6361/201220259}

\bibitem[{{Noterdaeme} {et~al.}(2012{\natexlab{b}}){Noterdaeme}, {Laursen},
  {Petitjean}, {Vergani}, {Maureira}, {Ledoux}, {Fynbo}, {L{\'o}pez}, \&
  {Srianand}}]{2012A&A...540A..63N}
{Noterdaeme}, P., {Laursen}, P., {Petitjean}, P., {et~al.} 2012{\natexlab{b}},
  \aap, 540, A63, \dodoi{10.1051/0004-6361/201118691}

\bibitem[{{Parks} {et~al.}(2018){Parks}, {Prochaska}, {Dong}, \&
  {Cai}}]{2018MNRAS.476.1151P}
{Parks}, D., {Prochaska}, J.~X., {Dong}, S., \& {Cai}, Z. 2018, \mnras, 476,
  1151, \dodoi{10.1093/mnras/sty196}

\bibitem[{{P{\'e}rez-R{\`a}fols} {et~al.}(2018){P{\'e}rez-R{\`a}fols},
  {Font-Ribera}, {Miralda-Escud{\'e}}, {Blomqvist}, {Bird}, {Busca}, {du Mas
  des Bourboux}, {Mas-Ribas}, {Noterdaeme}, {Petitjean}, {Rich}, \&
  {Schneider}}]{2018MNRAS.473.3019P}
{P{\'e}rez-R{\`a}fols}, I., {Font-Ribera}, A., {Miralda-Escud{\'e}}, J.,
  {et~al.} 2018, \mnras, 473, 3019, \dodoi{10.1093/mnras/stx2525}

\bibitem[{{P{\'e}roux} {et~al.}(2003){P{\'e}roux}, {McMahon},
  {Storrie-Lombardi}, \& {Irwin}}]{2003MNRAS.346.1103P}
{P{\'e}roux}, C., {McMahon}, R.~G., {Storrie-Lombardi}, L.~J., \& {Irwin},
  M.~J. 2003, \mnras, 346, 1103, \dodoi{10.1111/j.1365-2966.2003.07129.x}

\bibitem[{{Prochaska} \& {Herbert-Fort}(2004)}]{2004PASP..116..622P}
{Prochaska}, J.~X., \& {Herbert-Fort}, S. 2004, \pasp, 116, 622,
  \dodoi{10.1086/421985}

\bibitem[{{Prochaska} {et~al.}(2005){Prochaska}, {Herbert-Fort}, \&
  {Wolfe}}]{2005ApJ...635..123P}
{Prochaska}, J.~X., {Herbert-Fort}, S., \& {Wolfe}, A.~M. 2005, \apj, 635, 123,
  \dodoi{10.1086/497287}

\bibitem[{{Prochaska} {et~al.}(2015){Prochaska}, {O'Meara}, {Fumagalli},
  {Bernstein}, \& {Burles}}]{2015ApJS..221....2P}
{Prochaska}, J.~X., {O'Meara}, J.~M., {Fumagalli}, M., {Bernstein}, R.~A., \&
  {Burles}, S.~M. 2015, \apjs, 221, 2, \dodoi{10.1088/0067-0049/221/1/2}

\bibitem[{{Prochaska} \& {Wolfe}(1997)}]{1997ApJ...487...73P}
{Prochaska}, J.~X., \& {Wolfe}, A.~M. 1997, \apj, 487, 73,
  \dodoi{10.1086/304591}

\bibitem[{{Rahmati} {et~al.}(2014){Rahmati}, {Cravens}, {Larson}, {Fox},
  {Bougher}, {Lillis}, {Ledvina}, \& {Dunn}}]{2014AGUFM.P51B3932R}
{Rahmati}, A., {Cravens}, T., {Larson}, D.~E., {et~al.} 2014, in AGU Fall
  Meeting Abstracts, Vol. 2014, P51B--3932

\bibitem[{{Rauch}(1998)}]{1998ARA&A..36..267R}
{Rauch}, M. 1998, \araa, 36, 267, \dodoi{10.1146/annurev.astro.36.1.267}

\bibitem[{{Wolfe} {et~al.}(2005){Wolfe}, {Gawiser}, \&
  {Prochaska}}]{2005ARA&A..43..861W}
{Wolfe}, A.~M., {Gawiser}, E., \& {Prochaska}, J.~X. 2005, \araa, 43, 861,
  \dodoi{10.1146/annurev.astro.42.053102.133950}

\bibitem[{{Wolfe} {et~al.}(1986){Wolfe}, {Turnshek}, {Smith}, \&
  {Cohen}}]{1986ApJS...61..249W}
{Wolfe}, A.~M., {Turnshek}, D.~A., {Smith}, H.~E., \& {Cohen}, R.~D. 1986,
  \apjs, 61, 249, \dodoi{10.1086/191114}

\bibitem[{{Y{\`e}che} {et~al.}(2020){Y{\`e}che}, {Palanque-Delabrouille},
  {Claveau}, {Brooks}, {Chaussidon}, {Davis}, {Dawson}, {Dey}, {Duan},
  {Eftekharzadeh}, {Eisenstein}, {Gazta{\~n}aga}, {Kehoe}, {Landriau}, {Lang},
  {Levi}, {Meisner}, {Myers}, {Newman}, {Poppett}, {Prada}, {Raichoor},
  {Schlegel}, {Schubnell}, {Staten}, {Tarl{\'e}}, \&
  {Zhou}}]{2020RNAAS...4..179Y}
{Y{\`e}che}, C., {Palanque-Delabrouille}, N., {Claveau}, C.-A., {et~al.} 2020,
  Research Notes of the American Astronomical Society, 4, 179,
  \dodoi{10.3847/2515-5172/abc01a}

\bibitem[{{Zafar} {et~al.}(2013){Zafar}, {P{\'e}roux}, {Popping}, {Milliard},
  {Deharveng}, \& {Frank}}]{2013A&A...556A.141Z}
{Zafar}, T., {P{\'e}roux}, C., {Popping}, A., {et~al.} 2013, \aap, 556, A141,
  \dodoi{10.1051/0004-6361/201321154}

\end{thebibliography}
\bibliographystyle{aasjournal}

\end{document}